\documentclass[12pt]{iopart} %
\usepackage{url}

\RequirePackage[T1]{fontenc}

\RequirePackage{multirow} \RequirePackage{graphicx} \RequirePackage{mathptmx} %
\RequirePackage{flushend} \RequirePackage[numbers,sort&compress]{natbib} \RequirePackage[colorlinks,citecolor=blue,urlcolor=blue,linkcolor=blue]{hyperref}

\usepackage{graphicx} %
\usepackage{dcolumn} %
\usepackage{bm} %
\usepackage{amssymb} %
\usepackage{feynmf} %
\usepackage{hyperref} %
\usepackage{slashed} \usepackage{color} \usepackage{array} \usepackage{caption} \usepackage{lineno} \usepackage{subcaption} \expandafter\let\csname equation*\endcsname\relax \expandafter\let\csname endequation*\endcsname\relax \usepackage{amsmath} \usepackage{booktabs} \usepackage{multirow} \usepackage{subcaption} \usepackage{wrapfig} \usepackage{fancyhdr} \usepackage[section]{placeins}
\usepackage{float}
\usepackage{todonotes}
\usepackage[rightcaption]{sidecap}
\sidecaptionvpos{figure}{c}  %

\usepackage{parskip}
\newsavebox\mybox

\usepackage{acronym} \acrodef{PFlow}{particle flow}

\usepackage{cleveref}

\begin{document}

\pagenumbering{arabic}

\title{Prompting Particle Physics: Tokenized Multi-modal Foundation Models for Combinatorially Many Tasks}

\author{
Nilotpal Kakati$^{1,*}$\footnote[0]{$^*$Equal Contributions: \href{mailto:nilotpal.kakati@weizmann.ac.il}{nilotpal.kakati@weizmann.ac.il}, \href{mailto:daniel.thomas.murnane@cern.ch}{daniel.thomas.murnane@cern.ch} },
Daniel Murnane$^{2,3,*}$,  \\
Baran Hashemi$^4$, Samuel Klein$^5$, Jeffrey Krupa$^5$, \\
Eilam Gross$^1$, Lukas Heinrich$^6$, Michael Kagan$^5$}

\address{$^1$ Weizmann Institute of Science, Rehovot, Israel}
\address{$^2$ Niels Bohr Institute, University of Copenhagen, Copenhagen, Denmark}
\address{$^3$ Lawrence Berkeley National Laboratory, Berkeley, CA, USA}
\address{$^4$ Max Planck Institute for Mathematics in the Sciences, Leipzig, Germany}
\address{$^5$ SLAC National Accelerator Laboratory, Menlo Park, CA, USA}
\address{$^6$ Technical University of Munich, Munich, Germany}

\begin{abstract}
Reconstruction and simulation at a collider experiment are long chains of specialised algorithms, each tuned to a single step. We explore how one model can serve many of those steps at once, while still producing the intermediate objects (tracks, calorimeter cells, clusters, particles and jets) that make the chain interpretable; notably, this model operates on many types of data, from low-level sensor features to high-level physics features. To do so, we represent every object in a jet in a token vocabulary and train one model to map any subset of these modalities to any other. A task is then only a choice of which modalities to provide and which to request: particle flow, detector simulation and charged energy subtraction are all directions through the same set of weights. We train over all combinations of up to three input and three output modalities, with a decoder emitting tokens either autoregressively or in parallel. With tokenisation, both architectures train stably with little tuning. In evaluations, both models produce realistic reconstruction and simulation objects, with the autoregressive model particularly faithful to output from Geant4. The autoregressive model also outperforms a state-of-the-art particle flow algorithm on many typical jet reconstruction metrics.
\end{abstract}

\pagestyle{fancy}\fancyhf{}\fancyhead[R]{\thepage}\renewcommand{\headrulewidth}{0pt}

\vspace{1pc}

\section{Introduction}

Extracting physics from data recorded at the Large Hadron Collider (LHC) has historically involved a long chain of carefully tuned, physically motivated, and often heuristic stages. 
Data analyses start with raw energy deposits and build trajectories of charged particles~\cite{Fruhwirth1987Kalman, Strandlie2010TrackReco}, cluster the energy left in the calorimeters, and match these patterns to one another~\cite{Thomson2009PandoraPFA, ATLAS2017ParticleFlow, CMS2017ParticleFlow}. The chain then reconstructs the collision's decay products, i.e. leptons, photons, and hadrons, the latter of which is typically clustered into jets~\cite{Cacciari2008antikt, Cacciari2012FastJet}. These high-level representations of the event are used in analyses to constrain the parameters of the Standard Model, and search for New Physics extensions to it. High-fidelity simulators are employed to develop these algorithms and to compare data with theory. These simulators include modelling hard scattering, parton shower and hadronisation, and the passage of particles through the detector material and a digitised detector response~\cite{Sjostrand2015Pythia8, Agostinelli2003Geant4}. 

\begin{figure}[htbp]
 \centering
 \includegraphics[width=\linewidth]{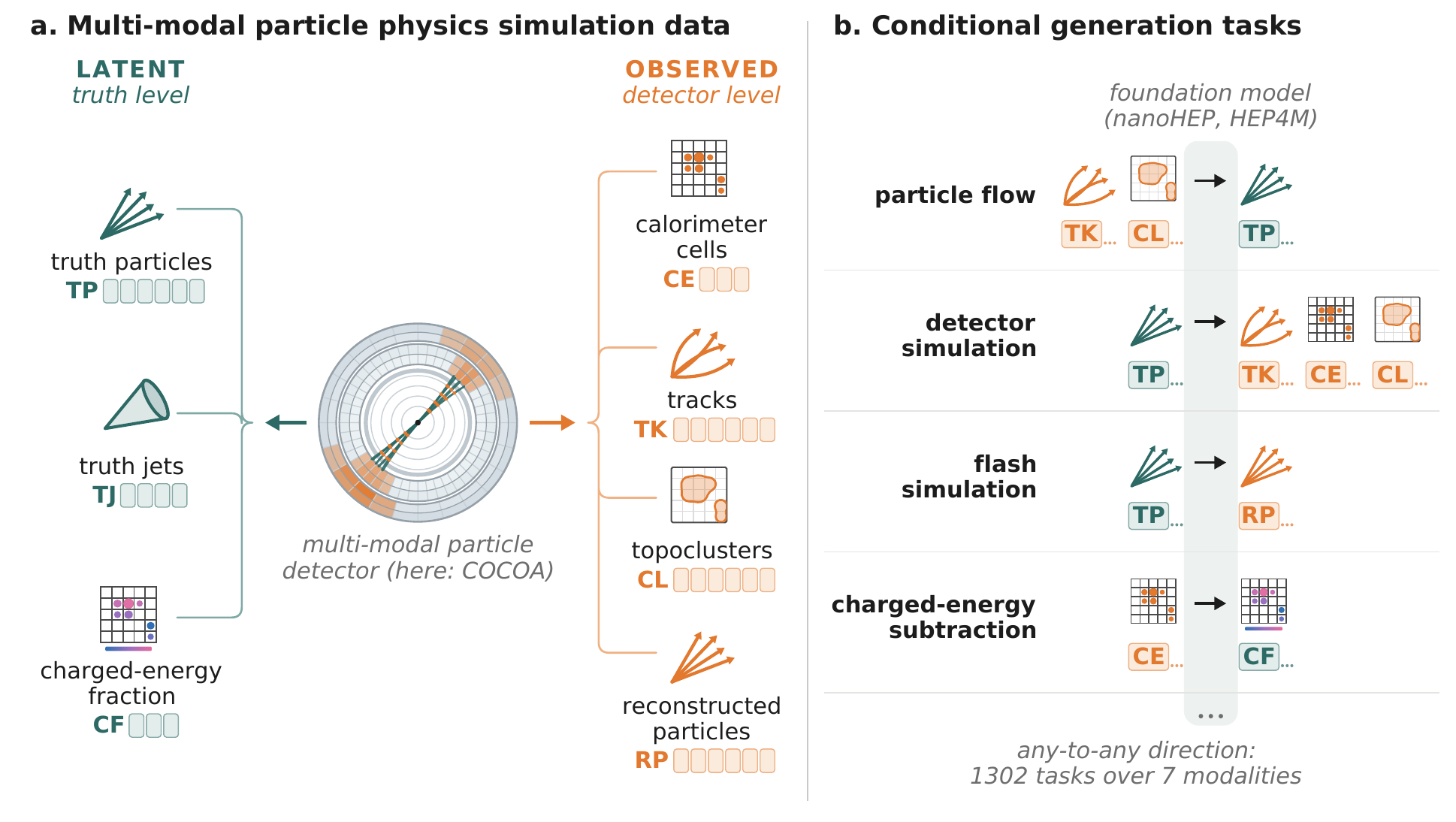}
 \caption{(a) The seven object modalities of a single jet, with the two-letter
 codes used in the figures of this paper. At truth level: the particles (TP), the
 jet (TJ) and the charged-energy fraction of each calorimeter cell (CF). At
 detector level: the calorimeter cells (CE), tracks (TK), topoclusters (CL) and
 the HGPflow particles reconstructed from them (RP). Each object is a short block
 of tokens.
 (b) A task is a choice of which modalities to provide and which to request.
 Four of the $1302$ directions are shown.}
 \label{fig:overview}
\end{figure}

In recent years, machine learning has been shown to solve individual links in these chains, in many cases matching the dedicated algorithms in accuracy or surpassing them in speed~\cite{Ju2021ExaTrkX, Pata2021MLPF, Kakati2024HGPflow, Paganini2017CaloGAN, Krause2021CaloFlow}. What has remained open is whether a \emph{single} model can replace several links at once, whilst remaining interpretable by producing intermediate objects (tracks, clusters, particles; Fig.~\ref{fig:overview}a). Such an approach would not aim to collapse the whole reconstruction into one opaque step and the intermediate objects can be inspected and calibrated in the usual way. This idea is closely related to developments in computer vision and natural language processing where general-purpose models~\cite{Mizrahi2023_4M, Radford2019GPT2} routinely consume raw measurements and produce predictions across many tasks. 

In this work,  we treat particle-physics reconstruction and simulation as a single multi-modal input-output problem. Every object in a jet has its continuous features converted to  discrete entries, or ``tokens'', in a codebook, combined across object types. The model is then trained to map any subset of these modalities to any other. Notably, these modalities cover a range of data types, from low-level sensor features to high-level physics features. Four such tasks are drawn in Fig.~\ref{fig:overview}b. Restricting each side to at most three of the seven modalities in this work leaves 1302 of them. With tokenised inputs and outputs, one generic model can perform any task defined over these objects simply by choosing which modalities to provide as input and which to request as output.

Tokenisation is an industry-standard approach to combining multiple modalities in a single model. In the DALL-E model~\cite{Ramesh2021DALLE}, images were tokenised so that text and pixels occupied a single sequence of inputs. 4M~\cite{Mizrahi2023_4M} extends the same recipe to many more modalities. Represented as tokens, every modality has a common input and output format, regardless of model backbone or task, and cross-entropy loss covers all combinations. The output is then a categorical distribution with stochastic outputs coming from sampling. Training remains simple classification, which shows particularly stable convergence behaviour. Using this approach, we investigate several multi-modal models that can perform a variety of reconstruction and simulation tasks simultaneously. Our contributions are:
\begin{enumerate}
    \item Two foundation models are presented: \textit{nanoHEP}, which decodes all output objects autoregressively, and \textit{HEP4M}, which predicts them in one pass with parallel token sampling. Both are trained down to sensor-level detector read-outs, capable of both reconstructing from and generating to this level of granularity.
    \item A systematic comparison of the computational and physics performance trade-offs between these two decoding extremes. The autoregressive nanoHEP is shown to outperform the state of the art in the benchmark task of particle flow reconstruction, and its simulated tracks and topoclusters match closely the behaviour of those from \textsc{Geant}4.
    \item The behaviour of multi-tasking and multi-modal physics models is explored across all $1302$ tasks with up to three input and three output modalities, to our knowledge the first evaluation of a physics model at this breadth, in particular to test whether they maintain performance on rarely seen tasks, and generalise to an unseen input combination.
\end{enumerate}

We begin with an overview of the two architectures and of how a task is expressed over a shared token dictionary (Sec.~\ref{sec:methods}). We then establish the performance of models trained towards single example tasks in each direction of the chain - reconstruction in Sec.~\ref{sec:headline} and simulation in Sec.~\ref{sec:generative}. The main result is the behaviour of the multi-task models across the task matrix and on unseen input combinations (Secs.~\ref{sec:mm_alltasks} and~\ref{sec:mm_transfer}).
 \section{Related Work} \label{sec:related}

The reconstruction and simulation of LHC events has been infused with machine learning to great effect over the last decade~\cite{Thais2022GNNReview}. For example, particle flow reconstruction can be recast as a learned graph or set prediction problem~\cite{Pata2021MLPF, Pata2023Scalable, DiBello2022HGPflow, Kakati2024HGPflow, PhysRevD.111.092015, Kobylianskii2025GLOW, CMS2026MLPF, Garcia2026EndToEnd}; charged-particle tracking as edge classification on a graph or as set prediction~\cite{Amrouche2019TrackML, Ju2021ExaTrkX, Caillou2022GNN4ITk, Lieret2023ObjectCondensation, VanStroud2024MaskFormerTracking}; jet flavour tagging as a classification over a jet's constituents~\cite{ATLASRNN2017,Hartman2020,Qu2019ParticleNet, Bols2020DeepJet, Mikuni2021PCT, Qu2022ParT, ATLASFTAG2022, ATLAS_GN2_2025, CMS_UParT_2024}; and detector simulation as a deep generative model of energy depositions~\cite{Paganini2017CaloGAN, Krause2021CaloFlow, Mikuni2022CaloScore, Buhmann2023CaloClouds, Hashemi2024IEAGAN, Amram2023CaloDiffusion, Kobylianskii2024CaloGraph, Hashemi2024DGMReview, Krause2024CaloChallenge,hildebrandt2026}.
Each of these examples is a single-stage model. A natural next step is to combine several stages at once. 
Every independent stage is a complex system to build, validate and keep in sync, and merging stages together shrinks that surface.
Furthermore, so-called ``end-to-end models'' or ``foundation models''\footnote{We here use these interchangeably, and delegate definitional work to \cite{Hallin2026FMreview}.} can propagate gradients across steps and ameliorate possible information bottlenecks at the boundaries of each stage.

A growing body of work has pushed in the end-to-end direction, which is usefully organised along three axes proposed in \cite{Hallin2026FMreview} that we adopt here: a) the pretraining task, b) the level of supervision and c) the amount of physics information the model is given. Pretraining objectives run from supervised classification~\cite{Qu2022ParT} through masked modelling~\cite{Golling2024MPM, Leigh2024MPMtokenization, Wildridge2024Bumblebee, Huang2024LanguageModelTracking}, next-token prediction~\cite{Birk2024OmniJet,birk2025omnijetalphac,Birk_2026}, contrastive objectives~\cite{Harris2024RS3L} and latent-space prediction~\cite{Katel2024JJEPA, Bardhan2025HEPJEPA}, to hybrids that combine generation with classification~\cite{Mikuni2024OmniLearn} and signature-oriented pre-training~\cite{Li2024Sophon}, and outside collider physics one backbone shared across images, spectra and scalars in astronomy~\cite{Parker2025AION1} and pre-training across particle imaging detectors~\cite{Young2026PandaDiplomacy}; models that incorporate symmetries~\cite{NEURIPS2024_277628cf,NEURIPS2025_29906cbd,lorentznet,bogatskiy2022pelican,Bogatskiy2022SymmetryReview} or differentiable physics~\cite{PhysRevD.110.052010}; several of these transfer to downstream tasks~\cite{Birk2024OmniJet, Mikuni2024OmniLearn, PhysRevD.111.092015,Mokhtar2026MLPFfoundation, Vigl2024Finetuning}.  Scaling models has also begun to be explored~\cite{ATL-SOFT-PUB-2026-002,vigl2026neural,amram2026neuralscalinglawsjet,uslu2026,bahl2026scalinglawsamplitudesurrogates}. Closest to the simulation directions of this work are Parnassus~\cite{Dreyer2024Parnassus, Dreyer2025ParnassusEvents} and FlowSim~\cite{Vaselli2024FlowSim}, generative models that produce reconstructed objects conditioned on generator-level particles for that one fixed direction. Complementary efforts include large open datasets that have begun to make foundation-model-scale~\cite{Bommasani2021FoundationModels} pre-training on collider data practical~\cite{Amram2024AspenOpenJets,jetset,Elitez2025ColliderML,Marshall2025ATLASOpenData,Moreno2025COLLIDE2V}.

In this work, we propose to build one model trained over all (input set, output set) pairs, up to three modalities on each side, across many modalities. Thus, reconstruction and simulation (and potentially mixed-modality tasks) become sequences predicted by one model, spanning from truth level information to detector hits and back. In this way, this work touches on all three axes, using pre-training strategies that can naturally use supervision when truth modalities are included, and integrating physics knowledge through the choice of modalities and tasks.

 \section{Methods} \label{sec:methods}

\subsection{Dataset and simulation} \label{subsec:dataset}

All events are produced with COCOA, the configurable calorimeter simulation for AI applications~\cite{DiBello2023COCOA}. COCOA generates hard-scatter events with Pythia\,8~\cite{Sjostrand2015Pythia8} and propagates the resulting particles through a Geant4~\cite{Agostinelli2003Geant4} model of a nearly hermetic detector: a segmented electromagnetic and hadronic calorimeter in a barrel and endcap arrangement, with an optional inner tracker of silicon and iron layers immersed in a magnetic field. From the raw energy deposits COCOA also runs a light layer of standard reconstruction - topological clustering of calorimeter cells and anti-$k_t$ jet finding~\cite{Cacciari2008antikt} - and records the truth particles from the generator. Tracks are provided as parameterised objects fast-simulated from the charged truth particles. This yields for every event a hierarchy of objects (Fig.~\ref{fig:overview}a): calorimeter cells, tracks, topoclusters, HGPflow~\cite{DiBello2022HGPflow, Kakati2024HGPflow} reconstructed particles, truth particles, truth jets, and the per-cell truth charged-energy fraction, seven modalities in all. The sample is single-jet events: $88.9$M for training and $100$k each for validation and test (\ref{app:dataset}). The truth-jet $p_T$ has a median of $46$ GeV ($8$ to $153$ GeV from the $5$th to the $95$th percentile), and $99\%$ of truth particles have $|\eta| < 2.5$. Table~\ref{tab:modalities} lists each modality with its features, its tokens and its mean number per event (\ref{app:modalities}).

\begin{table}[htbp]
 \centering
 \resizebox{\linewidth}{!}{%
 \begin{tabular}{llllcr}
  \toprule
  modality & code & content features (jointly quantised) & content tokens & position tokens & per event \\
  \midrule
  tracks & TK & $\sqrt{p_T}$, $d_0$ ($|d_0| \le 0.2$~mm), $z_0$ ($|z_0| \le 0.4$~mm) & $3 \times 256$ & $3$ & $3.6$ \\
  topoclusters & CL & $\sqrt{E}$, $\rho$, logit(EM fraction) & $3 \times 256$ & $3$ & $11.6$ \\
  truth particles & TP & $\sqrt{p_T}$, $\sqrt{E}$ & $3 \times 128$ & $3$ & $6.7$ \\
  HGPflow particles & RP & $\sqrt{p_T}$ & $3 \times 128$ & $3$ & $6.5$ \\
  truth jets & TJ & $\sqrt{p_T}$, $\sqrt{E}$ & $8 \times 128$ & $3$ per head & $1$ \\
  calorimeter cells & CE & $E^{0.3}$ per cell, $8 \times 8$ patches ($4 \times 4$ in the coarsest layer) & $3 \times 2048$ per patch & $3$ (nanoHEP) / none (HEP4M) & $156$ patches \\
  charged-energy fraction & CF & logit(charged energy fraction) per cell, same patches & $3 \times 2048$ per patch & $3$ (nanoHEP) / none (HEP4M) & $156$ patches \\
  \bottomrule
 \end{tabular}}
 \caption{The seven modalities, the two-letter codes used in the figures, and their tokenisers. Content tokens are written as $n_q \times K$ (residual levels $\times$ codebook size); the jet has eight heads of one token each. Position tokens are the $1024$-bin encodings of $(\eta, \cos\phi, \sin\phi)$; for a patch they are its binned centre, used by nanoHEP only, since in HEP4M a patch's location is its index in the fixed $156$-patch grid. The last column is the mean number of objects per event that the models see, within the window around the jet axis, on the test sample (on average $135$ of the at most $256$ stored cells are non-empty).}
 \label{tab:modalities}
\end{table}

\subsection{Tokenisation} \label{subsec:tok}

Every object in an event is represented as a short sequence of ``tokens'' drawn from a modality-specific ``codebook''~\cite{vandenOord2017VQVAE}. We tokenise all objects into a set of indices. One model with a categorical output then handles every modality. We split each object's attributes into two groups: content features are quantised to a learned codebook, while positions are binned in a fixed grid (\ref{app:implementation}). The \emph{content} attributes are quantised \emph{jointly} with a residual vector-quantised variational autoencoder (VQ-VAE)~\cite{vandenOord2017VQVAE, Lee2022RQVAE} (Table~\ref{tab:modalities}). A VQ-VAE encodes an object's attributes to a latent vector and replaces it by the nearest entry of a learned codebook; the index of that entry is the token, and a decoder trained alongside maps the entry back to the attributes. The residual variant stacks $n_q$ codebooks, each quantising the residual left by the previous, so an object's content becomes $n_q$ tokens that refine one another (\ref{app:implementation}, Fig.~\ref{fig:residual_vqvae}). The \emph{position} attributes $(\eta, \cos\phi, \sin\phi)$ are each binned into $1024$ uniform bins. The grid is the same for every modality, but the tokens are not: a track and a topocluster in the same $\eta$ bin receive different tokens. The azimuth is carried as the pair $(\cos\phi, \sin\phi)$ such that tokenisation is continuous across the $\phi = \pm\pi$ periodicity.

\begin{figure}[t]
\centering
\begin{subfigure}{\textwidth}
 \centering
 \includegraphics[width=0.92\linewidth]{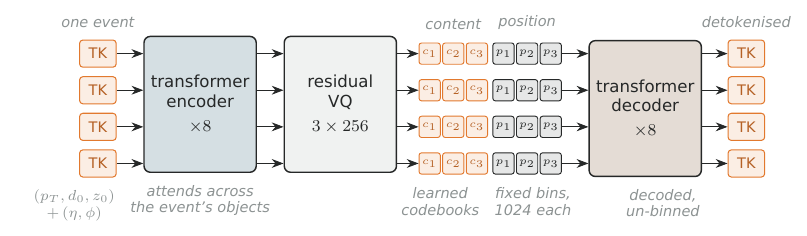}
 \caption{}
\end{subfigure}

\vspace{0.6em}
\begin{subfigure}{\textwidth}
 \centering
 \includegraphics[width=0.62\linewidth]{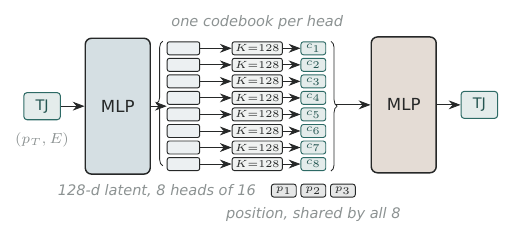}
 \caption{}
\end{subfigure}

\vspace{0.6em}
\begin{subfigure}{\textwidth}
 \centering
 \includegraphics[width=\linewidth]{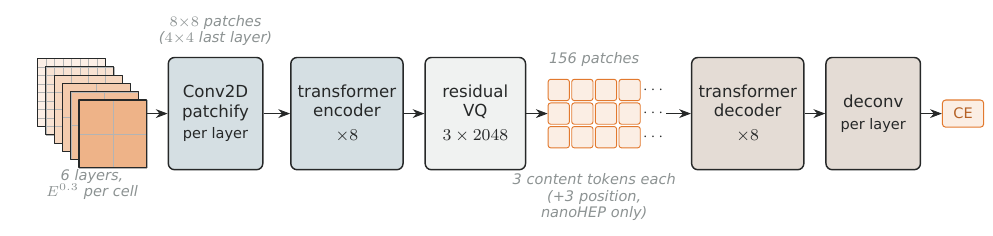}
 \caption{}
\end{subfigure}

\caption{The three tokeniser families. (a) Set-like objects (tracks shown; topoclusters and particles have the same form): each object becomes three residual content tokens and three binned position tokens (the residual cascade is drawn in Fig.~\ref{fig:residual_vqvae}). (b) The jet: a 128-dimensional latent split into eight heads, each matched to its own codebook; each head's index is paired with the jet's three position tokens. (c) The calorimeter-cell image (and the charged-energy fraction): six layers cut into $156$ patches of three tokens each, with no position tokens in HEP4M since a patch's location is its index (nanoHEP adds the binned patch centre).}
\label{fig:tokenizers}
\end{figure}

The mechanics of the residual quantiser and position-binning are detailed in \ref{app:implementation}. We here briefly sketch the most impactful parameter points. The codebooks are deliberately narrow - $8$-dimensional, projected down from the tokeniser's width of $128$ - which improves both codebook usage and reconstruction, behaviour that is also seen in image tokenisers~\cite{Yu2022ViTVQGAN}. They are initialised by $k$-means over a batch of encodings and thereafter updated by exponential moving average rather than by gradient, which keeps training stable when many codes are used rarely. Each modality is handled by its own tokeniser (Fig.~\ref{fig:tokenizers}). 

Calorimeter cells are tokenised specially: as an image rather than as a set. The cells of each of the six calorimeter layers within a fixed window around the jet axis are binned at the layer's hardware-level granularity, e.g. effectively forming multiple layers of jet images~\cite{jetimage2015,jetimage2016}, and the window is cut into square patches of $8 \times 8$ cells ($4 \times 4$ in the coarsest layer), $156$ patches per event. This patching strategy was inspired by the approach in vision transformers~\cite{dosovitskiy2021an}. Each patch is then embedded and quantised by the same residual VQ-VAE as the other modalities, with its centre binned as its position. The decoder returns the cell features and an occupancy flag per cell. 

Jets are global objects, one per event with no set structure to exploit, and are tokenised without a residual cascade. A bottleneck MLP encodes the content features into a 128-dimensional latent, which is split into eight 16-dimensional heads. Each head is matched against its own 128-entry codebook by scaled dot-product attention, so every jet receives eight content indices~\cite{Mama2021NWTTN}. Assignment is a straight-through Gumbel-softmax during training and an argmax at inference. Position is binned as for any other object.

\subsection{Tasks} \label{subsec:tasks}

\begin{figure}[t]
 \centering
 \includegraphics[width=\linewidth]{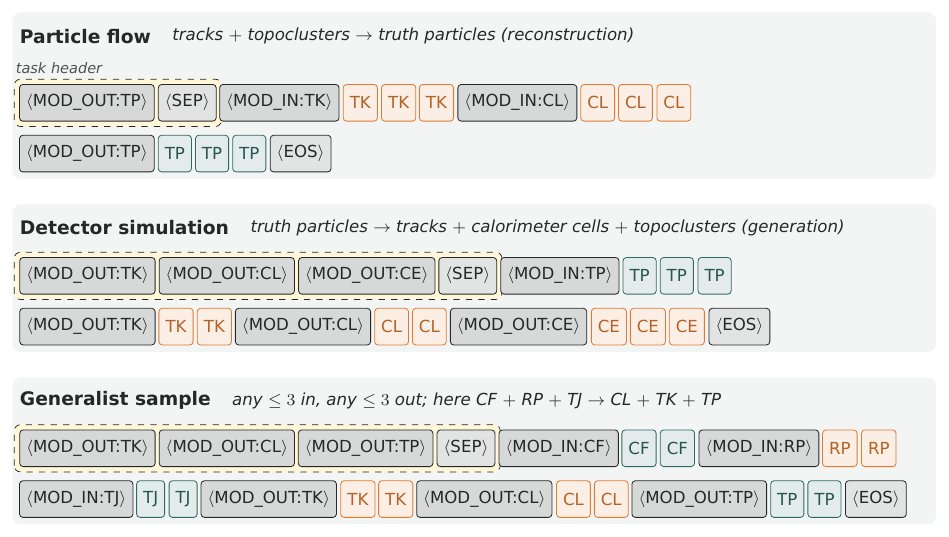}
 \caption{A task is an input/output modality split of the same tokenised
 event, written as one token sequence: a task header (shaded) of the requested
 output tags closed by the task separator, the input blocks, the output blocks, and
 end-of-sequence. Top: particle flow (tracks and topoclusters in, truth
 particles out). Middle: detector simulation (truth particles in; tracks,
 cells and topoclusters out). Bottom: one sample from the generalist sampler,
 which draws up to three input and up to three output modalities from the
 seven available ($1302$ distinct tasks); here truth cell fractions, HGPflow
 particles and jets in, and topoclusters, tracks and truth particles out.
 Coloured tokens are objects; grey tokens are the structural header, role
 tags, and end-of-sequence.}
 \label{fig:task_sequences}
\end{figure}

A major benefit of having a common token-based interface is that the core model is common to all tasks. As such, a task is completely defined by the modalities provided (the ``prompt'') and the modalities requested (the generated ``completion''). The same event and the same weights give particle flow~\cite{BUSKULIC1995481}, detector simulation, or flash simulation depending only on this split (Fig.~\ref{fig:task_sequences}). Particle flow is $\{\mathrm{TK}, \mathrm{CL}\} \rightarrow \{\mathrm{TP}\}$; detector simulation runs the sequence in reverse, $\{\mathrm{TP}\} \rightarrow \{\mathrm{TK}, \mathrm{CL}, \mathrm{CE}\}$; reconstruction refinement could be described as $\{\mathrm{RP}\} \rightarrow \{\mathrm{TP}\}$, and so on. Most previous models fix this split at design time, building it into the shapes of their inputs and loss. Leaving the split free makes every direction through the object hierarchy available to one set of weights, including directions not explicitly selected when the model was designed. The only ambiguities are how to request a modality and how to ensure the correct number of its objects are produced. We explore two ways of doing this, one for each of the autoregressive and parallel architectures, described in Sec.~\ref{subsec:fm_arch}.

To train the multi-task foundation models, at each step we sample up to three of the seven modalities as inputs and up to three of the remainder as outputs. Counting these gives $7\times 41 + 21\times 25 + 35\times 14 = 1302$. The sampler is uniform over them with one exception: the particle flow direction ($\{\mathrm{TK}, \mathrm{CL}\} \rightarrow \{\mathrm{TP}\}$) is over-sampled, to ensure that approximately $10\%$ of task training samples are exactly the particle flow task, to improve training stability and to study the impact of a dominant task. Note however that particle flow reconstructions are short sequences, and so only around $2\%$ of compute is dedicated to this task. A model trained in this way learns all of the directions jointly (Sec.~\ref{sec:multimodal}). Since the sampler never draws more than three inputs, four-input directions are never presented during training. We use this gap in Sec.~\ref{sec:generalization} to test whether the model can beneficially compose modality combinations it has not seen before.

\subsection{Model architectures} \label{subsec:fm_arch}

We explore two methods for how a tokenised sequence can be produced: emit the output tokens autoregressively, or all tokens in parallel. For both, all input tokens are processed jointly as context to drive decoding. Autoregression is the natural choice for a sequence of tokens. Parallel decoding can accelerate prediction at the cost of a slight mismodelling of the joint distribution.

The two architectures consume an identical frozen tokenisation and make per-token categorical predictions with cross-entropy loss. They also use the same temperature and top-$k$ sampling and the same frozen VQ decoders for detokenisation (converting tokens back to physical space). In both cases, objects are ordered by decreasing $p_T$ (track, jet, particle) or energy (cluster, cell) within each modality. Also in both cases, within each object, the $n_q$ residual tokens are decoded conditionally. I.e. for the $n^{th}$ track, the sampled token at one residual level is embedded and used to conditionally generate the next, to align with the RVQ-VAE approach. In training, this is achieved with teacher forcing~\cite{10.1162/neco.1989.1.2.270}.

\subsubsection{Iterative (autoregressive) decoding} \label{subsubsec:autoregressive}

\begin{figure}[t]
 \centering
 \includegraphics[width=0.62\linewidth]{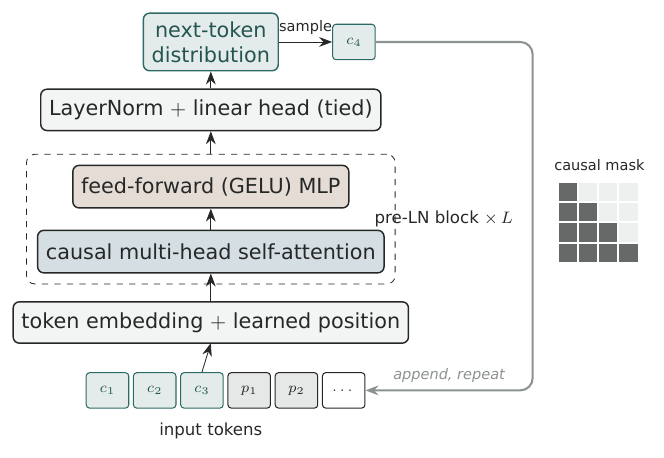}
 \caption{Autoregressive (iterative) decoding model: a decoder-only GPT-2-class
 transformer over the shared token vocabulary. Each token attends causally to
 all previous tokens and the next token is sampled from the predicted
 distribution; a structural mask constrains which tokens are legal at each
 step, and a key-value cache avoids recomputation during decoding.}
 \label{fig:autoregressive_model}
\end{figure}

The autoregressive model, \textit{nanoHEP}, is a decoder-only transformer~\cite{Vaswani2017Transformer} of the GPT-2 family~\cite{Radford2019GPT2}, taken broadly unchanged from the nanoGPT~\cite{Karpathy_nanoGPT} implementation. The approach is similar to prior work on token-based autoregressive models of jets and calorimeter showers~\cite{Birk2024OmniJet,birk2025omnijetalphac,Birk_2026}: tokens as input and learned positional embeddings, a stack of pre-LayerNorm~\cite{Xiong2020PreLN} blocks each with causal multi-head self-attention and a two-layer GELU~\cite{Hendrycks2016GELU} feed-forward network, and a final linear head, tied to the input embedding~\cite{Press2017WeightTying}, that predicts the next token (Fig.~\ref{fig:autoregressive_model}). Causal masking means every token attends only to those before it, so the model factorises the joint distribution over the whole sequence into a product of next-token conditionals.

\begin{equation}
q_{\theta}(y \mid x) = \prod_{t=1}^{|y|} q_{\theta}\big(y_t \mid x, y_{<t}\big).
\label{eq:ar_factorisation}
\end{equation}

The nanoHEP model uses one vocabulary for everything. Each modality's content codebooks and position bins are assigned their own range of integer IDs and the ranges are concatenated, so each id belongs to exactly one modality and one value. The eight jet heads share one range of $128$ ids, and a head is identified by its position in the jet. For the seven modalities this is $36{,}224$ ids. A further $2M+3 = 17$ IDs are structural for $M$ modalities: an input tag $\langle\mathrm{MOD\_IN}{:}m\rangle$ and an output tag $\langle\mathrm{MOD\_OUT}{:}m\rangle$ per modality, a task separator, an end-of-sequence token, and padding, for $36{,}241$ IDs in total. One embedding table and its weight-tied softmax cover the whole vocabulary.

In nanoHEP the token sequence is pre-formatted as follows: A header lists the requested output modalities in canonical order and is closed by the task separator. The input region follows, with the objects of each provided modality grouped after its input tag. The output region contains the generated objects, grouped after their output tags, and closes with end-of-sequence tokens (``EOS"). During training, the loss is applied from the first output tag in the body. We note that alternative encodings of the task (or prompt) may be possible but leave a detailed study for future work. 

At object boundaries the model chooses whether to emit another object of the same modality, an output tag opening the next modality, or end-of-sequence. Modality choice, cardinality, and completion are therefore the same learned next-token decision. A structural mask restricts it to the options legal under the sequence grammar - if a model has chosen to output a stream of tracks, it cannot output a cell until it switches modality. Relative to nanoGPT we make no changes to the transformer core; our additions are all at the interface - the union multimodal vocabulary, the structural mask, and a key-value cache~\cite{Shazeer2019MQA, Pope2023TransformerInference} for fast decoding (\ref{subsec:training_inference}).

\subsubsection{Parallel decoding} \label{subsubsec:parallel}

The parallel decoding model, \textit{HEP4M}, is an adaptation of the 4M architecture~\cite{Mizrahi2023_4M}, and benefits from prior work on token-based masked models applied to jets~\cite{Golling2024MPM, Leigh2024MPMtokenization}. To ensure the same dimensionality of all inputs, the tokens for each modality are summed with a learned per-modality vector (similar to positional encoding). As such, the modalities are unified in the embedding space rather than by offsetting their token ids, as in nanoHEP.

\begin{figure}[t]
 \centering
 \includegraphics[width=0.95\linewidth]{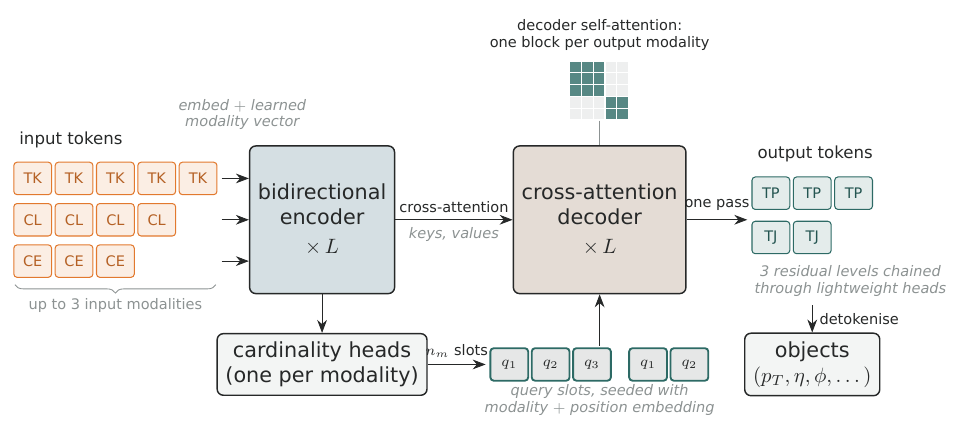}
 \caption{Parallel-decoding model. Input-modality token blocks are embedded
 and read by a bidirectional encoder; per-modality cardinality heads set the
 number of output slots; decoder query slots (seeded with modality and
 position embeddings) cross-attend to the encoder and predict all output
 tokens in one pass, which are detokenised back to objects.}
 \label{fig:parallel_model}
\end{figure}

HEP4M is an encoder-decoder transformer. Input tokens are provided grouped by object and modality and read by a bidirectional transformer encoder. Instead of a task header, a set of output modalities is requested by attaching per-modality cardinality heads to the model's final layer. For each modality, its head fixes the number $N$ of objects and $N$ decoder query slots are allocated, seeded with that modality's embedding and a position embedding. A block-diagonal self-attention mask separates the groups of slots. The decoder cross-attends to the encoder, predicts all output token distributions in a single forward pass, and routes each group to its own per-modality prediction head before sampling (Fig.~\ref{fig:parallel_model}). A slot therefore cannot emit the wrong modality: its modality is fixed when it is allocated.

Each object's sampled tokens never condition on another's. Within a sequence of residual tokens describing an element of a modality, however, the residual levels are decoded in order, with each sampled residual code conditioning the next. Relative to the original 4M we also handle position tokens with a separate binned-content output, and use RMS-normalised~\cite{Zhang2019RMSNorm} attention with cross-attention preceding self-attention in the decoder. The cardinality head is categorical: it predicts a distribution over $0, 1, \dots, N_{\max}$ and is sampled at inference time (for alternatives, we refer the reader to \ref{app:implementation}).

With $x$ and $y$ as in Eq.~\ref{eq:ar_factorisation}, the per-modality cardinality heads (\ref{app:cardinality_head}) first fix the counts $n$, and each object is then drawn independently:
\begin{equation}
q_{\theta}(y, n \mid x) = \prod_{m} q_{\theta}(n_m \mid x) \prod_{i=1}^{n_m} q_{\theta}\big(y^{m}_{i} \mid x, n\big),
\label{eq:parallel_factorisation}
\end{equation}
where $y^m_i$ is the $i$-th object of modality $m$, itself a block of tokens whose residual levels are decoded in dependency order (\ref{app:implementation} gives the expanded form).

\subsection{Evaluation metrics} \label{sec:metrics}

We propose three classes of metrics by which to evaluate the models discussed above. Benchmark metrics are evaluated on a fixed 100k-event test set, to ensure comparable statistics across models; the all-direction sweep of Sec.~\ref{sec:mm_alltasks} uses a cheaper protocol (the fast judge) described there. 

First, we report aggregated response and resolution for high-level objects. For example, for a reconstructed or generated jet we take the response $p_T^{\mathrm{pred}}/p_T^{\mathrm{truth}}$ and quote its median as the bias and its interquartile range ($Q_{75}-Q_{25}$) as the resolution. We use the IQR rather than a fitted width because it is robust to tails in the distribution.

Second, we report per-object marginals: the distributions of each object feature over the dataset, compared with the truth distributions by the Kolmogorov-Smirnov statistic. Objects are ordered by $p_T$, which lets the leading object of a prediction be paired with the leading truth object by rank alone; beyond the leading object, rank pairing is unreliable once resolution smearing reorders neighbouring objects, so we do not report matched per-object residuals.

Finally, we report two-sample tests (C2STs)~\cite{LopezPaz2017C2ST}: we train a classifier to separate a method's predicted objects from the \textsc{Geant}4 or reconstructed truth and report AUC on held-out validation objects. An AUC of $0.5$ means this classifier cannot separate the two; any excess is a single, comparable measure of mismodelling. The classifier also receives the event's inputs, as in the input-conditioned test of~\cite{Vaselli2024FlowSim}. It therefore tests whether the prediction is consistent with that event. The specific architecture used for the C2ST is a set transformer~\cite{Lee2019SetTransformer} that attends over the full object set, combined with summary features computed on both inputs and outputs.\footnote{The set transformer has two layers, four heads and width $48$, and is seeded with per-event summary features of the inputs and outputs. AUCs are averaged over five classifier retrainings on 100k events, with a fixed optimisation budget of at most $35$k steps and an event-wise $70/30$ train/test split; every campaign also runs a control with identical samples on both sides, which reads between $0.500$ and $0.502$.} A classifier of this kind gives a lower bound: with a stronger classifier or more events, the measured separation may grow. Quoted values are therefore comparable within this protocol rather than absolute~\cite{Das2024Limitations, Lee2024ConditionalTwoSample, Chatterjee2024KernelConditional}. In HEP4M, $\cos\phi$ and $\sin\phi$ are sampled in parallel, so nothing guarantees $\cos^2\phi + \sin^2\phi = 1$. We therefore renormalise these outputs onto the unit circle, and all HEP4M results are reported with this correction. nanoHEP decodes each token conditioned on those already sampled, so it can (and does) learn to satisfy the identity: under sampling, $\cos^2\phi + \sin^2\phi = 1.000 \pm 0.011$ for its tracks and $1.000 \pm 0.006$ for its topoclusters, against $0.97 \pm 0.21$ and $0.97 \pm 0.30$ for HEP4M before the correction (\ref{app:decoding}).

\section{Results} \label{sec:results} \label{sec:multimodal}

In this section we evaluate the two architectures of Sec.~\ref{sec:methods}, first on two standard tasks, particle flow reconstruction and detector simulation (Secs.~\ref{sec:headline} and \ref{sec:generative}), and then across every direction of the task matrix (Sec.~\ref{sec:mm_alltasks}). We additionally study task transfer (Sec.~\ref{sec:mm_transfer}) and report limitations (Sec.~\ref{sec:mm_limits}).

NanoHEP-pflow and HEP4M-pflow are trained on particle flow alone, nanoHEP-sim and HEP4M-sim on detector simulation alone. In contrast, nanoHEP-multi and HEP4M-multi are instead trained on the whole task matrix at once. The model sizes are as follows: nanoHEP-pflow and nanoHEP-sim have $94$M parameters and nanoHEP-multi $115$M, while HEP4M-pflow and HEP4M-sim have $114$M, and HEP4M-multi $133$M. Parameter counts include all trained weights and exclude the frozen tokenisers.

For the classifier-based two-sample tests we use two different classifiers: the benchmark tasks (particle flow and detector simulation) are judged with a high-precision classifier trained on $100$k events. Since applying this test to all $1302$ directions is computationally expensive, we train coarser classifiers (a ``fast judge''): $512$ events per direction ($128$ on the $707$ directions that output cells or charged-energy fractions) and thirty training epochs. Its scores are meaningful only relative to one another: with so few events it compresses differences towards $0.5$, so a score near $0.5$ means the fast judge does not separate the samples, not that they are indistinguishable from truth. Across the matrix (Sec.~\ref{sec:mm_alltasks}) every direction is decoded the way the generative benchmark is decoded in Sec.~\ref{sec:generative}, by sampling at temperature $1$. The transfer study of Sec.~\ref{sec:mm_transfer} compares checkpoints with argmax decoding on the reconstruction directions and sampling at temperature $1$ on the generation directions.

\subsection{The Particle Flow Task} \label{sec:headline} \label{sec:mm_anchors}

\begin{figure}[htbp]
 \centering
 \includegraphics[width=\linewidth]{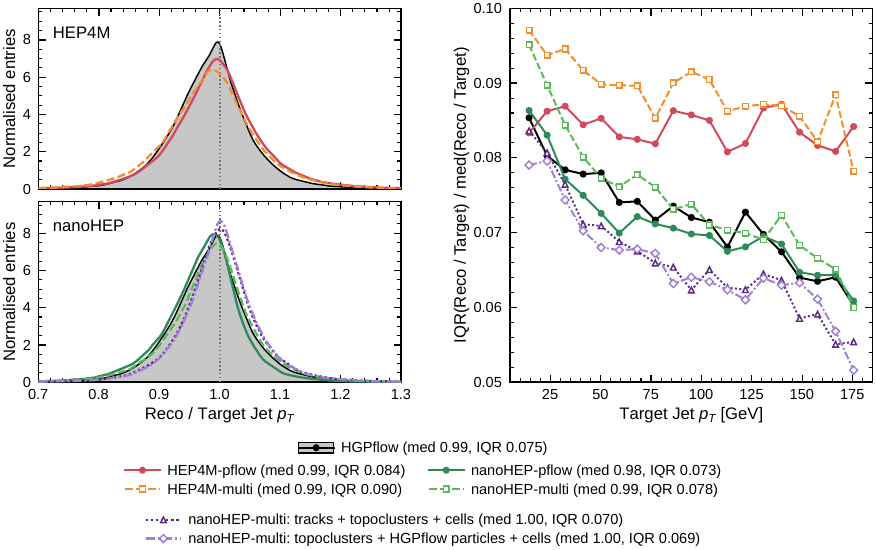}
 \caption{Jet-level agreement on particle flow, for the single-task models (nanoHEP-pflow, HEP4M-pflow), the multi-task models (nanoHEP-multi, HEP4M-multi) and HGPflow: the reconstructed-to-target jet $p_T$ response (left; HEP4M models above, nanoHEP models below, each against HGPflow as the filled distribution; smoothed with a Gaussian kernel of width $0.0056$, the same for every curve) and its IQR/median in bins of target jet $p_T$ (right). The legend gives the median and IQR of the response. Two further nanoHEP-multi curves (purple) use the extra inputs of Table~\ref{tab:input_ladder}: tracks, topoclusters and cells, and topoclusters, HGPflow particles and cells, with no further training.}
 \label{fig:headline_jets}
\end{figure}

The particle flow task aims to predict a variable-length list of particles and their respective properties, from a fixed set of tracks and topoclusters. In a typical HEP workflow, these particle flow objects are then used to isolate single particles or clustered as jets. We use the state-of-the-art ML-based particle flow model HGPflow~\cite{DiBello2022HGPflow, Kakati2024HGPflow} as a baseline for comparison.

\begin{SCtable}[1.5][htbp]
 \centering
 \small
 \begin{tabular}{lccc}
  \toprule
  & \multicolumn{2}{c}{jet $p_T$ response} & classifier \\
  \cmidrule(lr){2-3}
  model & median & IQR & AUC \\
  \midrule
  HGPflow                           & $0.986$ & $0.0754$ & $0.902$ \\
  nanoHEP-pflow                     & $0.978$ & \textbf{0.0733} & \textbf{0.691} \\
  nanoHEP-pflow & \multirow{2}*{$0.982$} & \multirow{2}*{$0.0758$} & \multirow{2}*{$0.697$} \\
  \ \ \ (matched, 29k steps) & & &  \vspace{0.1cm}\\
  nanoHEP-multi                     & $0.990$ & $0.0781$ & $0.703$ \\
  HEP4M-pflow                       & \textbf{0.994} & $0.0836$ & $0.940$ \\
  HEP4M-multi                       & $0.986$ & $0.0898$ & $0.939$ \\
  \bottomrule
 \end{tabular}
 \caption{Particle flow reconstruction performance, using argmax decoding. We report the jet $p_T$ response median and IQR, and the classifier AUC of Sec.~\ref{sec:metrics}. The best value in each column is in bold, and all measures are in Table~\ref{tab:pflow_measures}.}
 \label{tab:pflow_main}
\end{SCtable}

Every result in this section is evaluated on approximately $100$k test events. We evaluate with the three metric classes of Sec.~\ref{sec:metrics}~\footnote{To our knowledge, evaluating particle flow \emph{reconstruction} with a learned classifier is novel. The particle flow literature conventionally assesses jet response and resolution, particle efficiency, and fake-rate metrics~\cite{Thomson2009PandoraPFA, Pata2021MLPF, DiBello2022HGPflow, Pata2023Scalable}}. As high-level aggregate object we use jets and report the $p_T$ response median (bias) and IQR (resolution), and the angular distance between the reconstructed and target jets (Fig.~\ref{fig:headline_jets}; angles in \ref{app:tasks}, Fig.~\ref{fig:headline_angles}). The jet-level numbers are computed against the raw truth particles.

We observe that both models, despite being trained with no physics constraints or inductive biases, are close at jet level to the state-of-the-art reconstruction algorithm (Fig.~\ref{fig:headline_jets} left). nanoHEP in particular outperforms HGPflow in reconstructing the jet momentum both in resolution level as well as by classifier score. The single-task HEP4M-pflow model achieves the best calibrated response with its median closest to unity. We also report the jet resolution as a function of the jet momentum $p_T$ (Fig.~\ref{fig:headline_jets} right). The results are summarized in Table~\ref{tab:pflow_main}. nanoHEP-pflow is therefore the best model on the particle flow task (tracks and topoclusters to particles). For particle reconstruction from other inputs, nanoHEP-multi does better still, reaching an IQR of $0.069$ and an AUC of $0.66$ from four inputs (Table~\ref{tab:input_ladder}). Trained from scratch on particle flow alone for the same $29$k particle flow steps that nanoHEP-multi received, nanoHEP reaches $0.0758$, with an AUC of $0.697$ (Table~\ref{tab:pflow_main}, ``matched'').

We report the per-particle marginals in the appendix (Fig.~\ref{fig:mm_pflow_marg}). To investigate the performance difference between autoregression (nanoHEP) and parallel decoding (HEP4M) more carefully, we also include other particle-cloud-level measures of fidelity. The Energy-Mover's Distance (EMD) allows a per-event measure of particle flow accuracy, as does a compatibility judge model - background for these measures is provided in \ref{app:pflow_measures}. The behaviour seen at jet level remains: HGPflow and nanoHEP-pflow are within $0.002$ in jet IQR and $1\%$ in EMD. However, nanoHEP-pflow is also much less separable from the truth (AUC $0.69$ against $0.90$; Table~\ref{tab:pflow_main}).

While topoclusters and tracks are the traditional inputs to the particle flow task, we observe that adding more modalities that either add more lower level of information (e.g. calorimeter cells) or preprocess the data (e.g. HGPflow particles) can further boost performance. These additional tasks are also shown in Fig.~\ref{fig:headline_jets}. We discuss this in more detail in Sec.~\ref{sec:generalization}.

\begin{SCfigure}[30][htbp]
 \centering
 \includegraphics[width=0.55\linewidth]{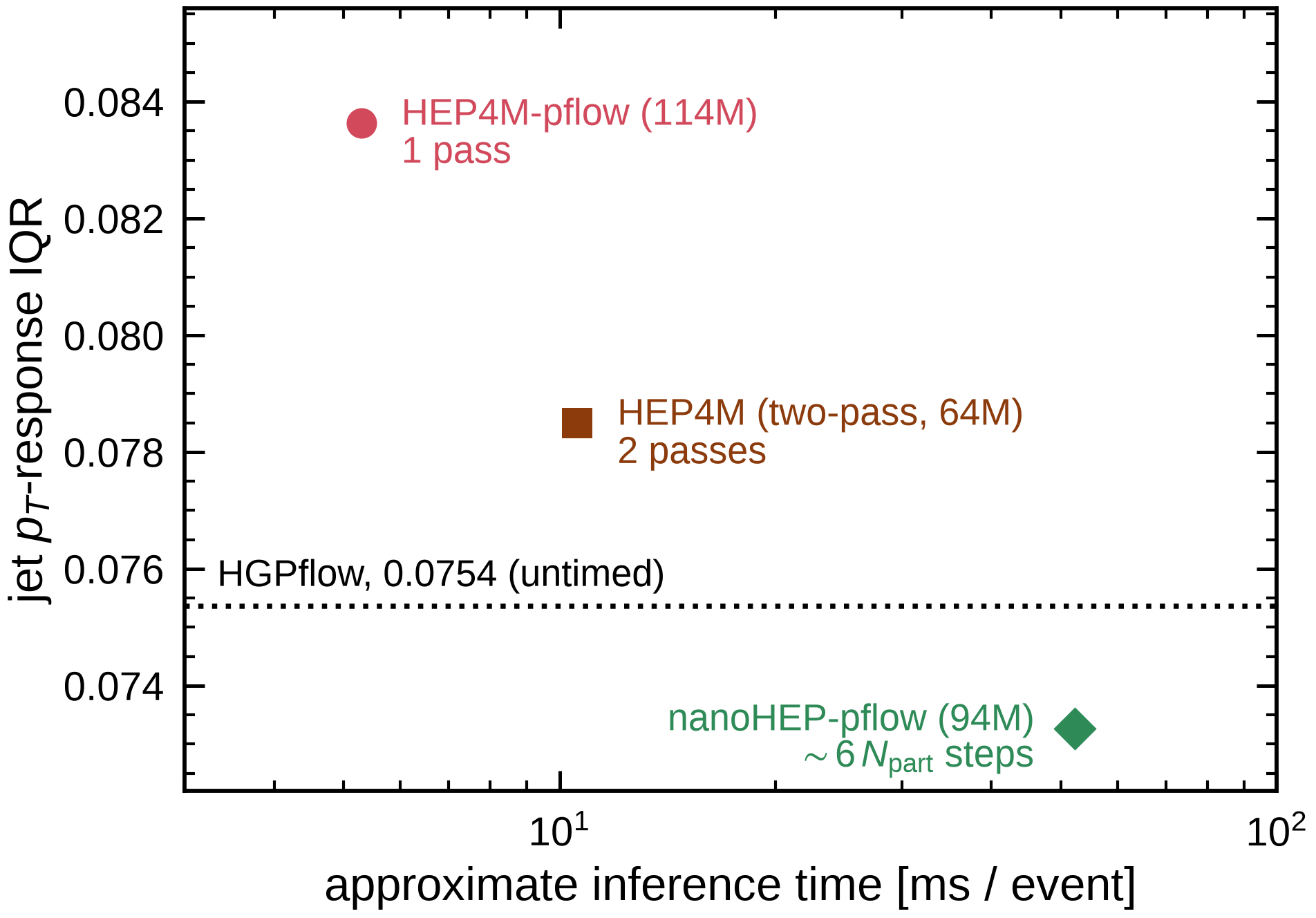}
 \caption{Jet $p_T$-response IQR versus approximate inference time per event for HEP4M-pflow, HEP4M (two-pass) of \ref{subsec:cond_twopass}, and nanoHEP-pflow. The dotted line is the HGPflow IQR, which is not timed on this axis.}
 \label{fig:headline_pareto}
\end{SCfigure}

\subsubsection{Computational performance}

The high-quality reconstruction of nanoHEP-pflow must be considered in tandem with its cost, as it decodes iteratively compared to the single forward pass of HEP4M\footnote{The three residual levels of each object are chained through lightweight prediction heads on the fixed decoder output, not through further decoder passes; Sec.~\ref{subsec:fm_arch}}. Fig.~\ref{fig:headline_pareto} makes the trade-off explicit. To close the gap, we investigate a partial autoregression in HEP4M, where the leading particle is generated individually while all remaining ones are generated in parallel. We observe a significant improvement at modest additional computational cost. The emerging Pareto frontier between inference cost and fidelity is shown in Fig.~\ref{fig:headline_pareto} and a full study is detailed in \ref{subsec:cond_twopass}.

\subsection{The Simulation Task} \label{sec:generative}

We turn from reconstruction to generation. In the tokenized framework this is only another choice of prompt: truth particle $\rightarrow$ track $+$ topocluster. In COCOA, tracks are fast-simulated from the charged truth particles; for brevity we refer to the full simulated target as \textsc{Geant}4. We report the same three metric classes: classifier two-sample tests are shown in Table~\ref{tab:sim_main}. For tracks, the tokeniser round trip alone is separable from raw \textsc{Geant}4 at an AUC of $0.54$, from the $d_0$ and $z_0$ range, so nanoHEP-sim's $0.54$ is at the level the tokenisation allows. Per-event aggregate measures and per-object marginals are listed in \ref{app:decoding}. Similarly to the particle flow task, the autoregressive nanoHEP model performs better than the HEP4M model at the cost of increased inference time. 

\begin{figure}[htbp]
 \centering
 \includegraphics[width=\linewidth]{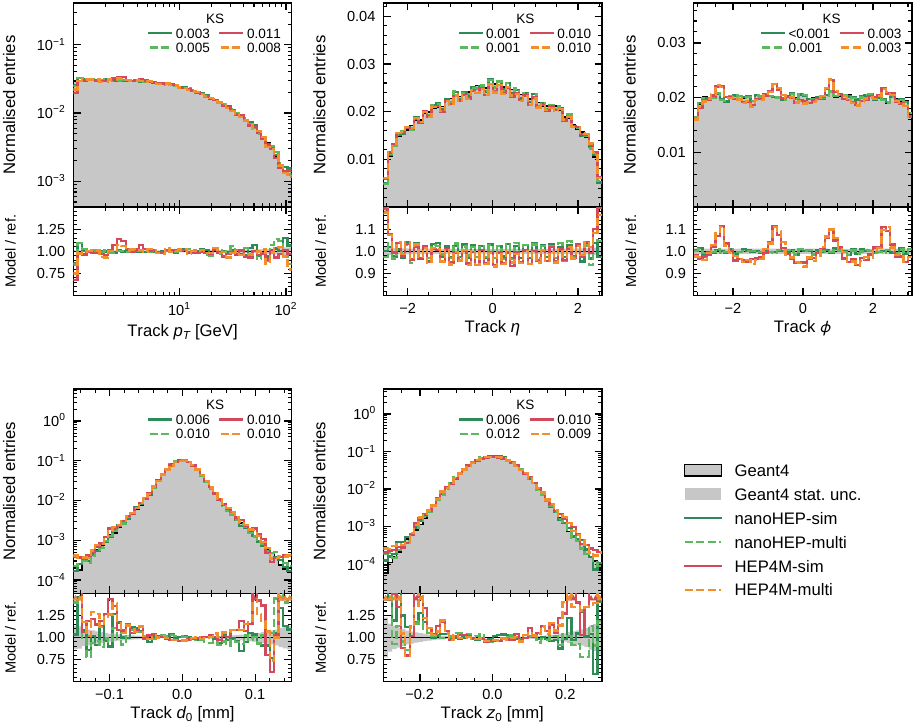}
 \caption{Generated track $p_T$, $\eta$, $\phi$, $d_0$ and $z_0$ for the single-task (solid) and multi-task (dashed) models, sampled at $T=1$, against \textsc{Geant}4. The KS distance to \textsc{Geant}4 is given for each model.}
 \label{fig:gen_track}
\end{figure}

\begin{figure}[htbp]
 \centering
 \includegraphics[width=\linewidth]{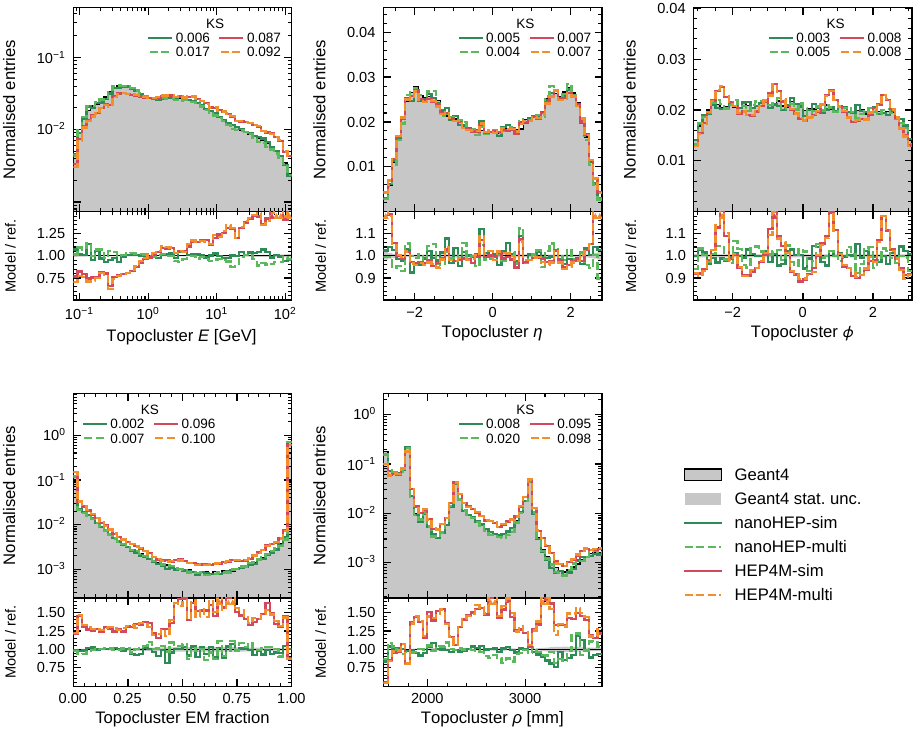}
 \caption{Generated topocluster $E$, $\eta$, $\phi$, EM fraction and $\rho$ for the single-task (solid) and multi-task (dashed) models, sampled at $T=1$, against \textsc{Geant}4. The KS distance to \textsc{Geant}4 is given for each model.}
 \label{fig:gen_topo}
\end{figure}

Given the fully sampled configuration, the single-object distributions from both nanoHEP-sim and HEP4M-sim align closely with those of full simulation (Figs.~\ref{fig:gen_track} and \ref{fig:gen_topo}). The track tokeniser covers $|d_0| \le 0.2$~mm and $|z_0| \le 0.4$~mm (Table~\ref{tab:modalities}). The $1.3\%$ of \textsc{Geant}4 tracks outside this range are mapped to the edge codes; the plotted $d_0$ and $z_0$ ranges in Fig.~\ref{fig:gen_track} exclude these edge bins. HEP4M samples $\cos\phi$ and $\sin\phi$ independently, which pulls some angles towards the diagonals ($\pm\pi/4$, $\pm 3\pi/4$) in Figs.~\ref{fig:gen_track} and~\ref{fig:gen_topo}. Cardinality remains a difficult problem. Both models return the exact track count, but for topoclusters only nanoHEP-sim reproduces the multiplicity distribution; HEP4M-sim is too narrow, under-producing the high-multiplicity events, and the per-event residual is broad for both (Fig.~\ref{fig:gen_card}). A variety of approaches were tested with both models to alleviate this - such as indicator losses and cardinality regression - but no single approach provided the clear solution.

\begin{SCtable}[1.2][htbp]
 \centering
 \small
 \begin{tabular}{lcc}
  \toprule
  model & tracks & topoclusters \\
  \midrule
  nanoHEP-sim   & \textbf{0.541} & \textbf{0.517} \\
  nanoHEP-multi & $0.554$ & $0.525$ \\
  HEP4M-sim     & $0.972$ & $0.984$ \\
  HEP4M-multi   & $0.987$ & $0.985$ \\
  \bottomrule
 \end{tabular}
 \caption{AUC of classifier two-sample tests for detector simulation (truth particles $\rightarrow$ tracks and topoclusters) separating generated tracks or topoclusters from Geant4's, conditioned on the event's truth particles. Decoded by sampling at temperature $1$. Mean of five retrainings shown, with a spread of at most $0.007$. 
 Best value in each column in bold.}
 \label{tab:sim_main}
\end{SCtable}

\subsection{All tasks} \label{sec:mm_alltasks} 

\begin{figure}[htbp]
 \centering
 \includegraphics[width=\linewidth]{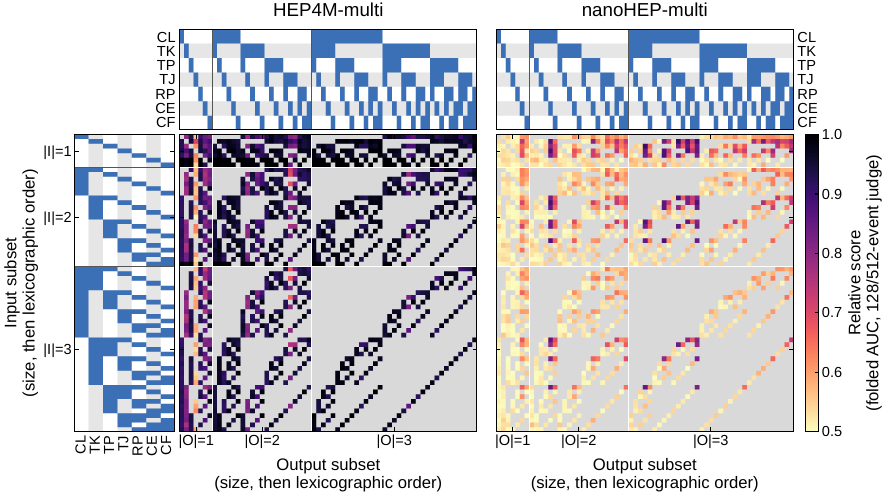}
 \caption{All $1302$ task directions for HEP4M-multi (left) and nanoHEP-multi (right), on the grid of input subsets (rows) versus output subsets (columns): the fast-judge score (Sec.~\ref{sec:mm_alltasks}; compare values only within this figure). Grey cells show undefined tasks, since output should not contain an input modality. The multiplicity and marginal distances for the same directions are in Fig.~\ref{fig:mm_alltasks_mult_ks}.}
 \label{fig:mm_alltasks}
\end{figure}

We run inference with nanoHEP-multi and HEP4M-multi on each of the $1302$ directions and score the outputs on three quantities: multiplicity, marginal agreement, and the fast-judge score. Further directions are in \ref{app:tasks}: reconstruction refinement (\ref{app:unfold}), flash simulation (\ref{app:flashsim}), jet simulation (\ref{app:jetsim}), charged-energy subtraction (\ref{app:chesub}) and topoclustering (\ref{app:topoclus}). Data and model scaling of nanoHEP-pflow are in \ref{app:scaling}. Fig.~\ref{fig:mm_alltasks} arranges the directions by input (rows) and output (columns) subset and shows the fast-judge score; the multiplicity and marginal distances are in Fig.~\ref{fig:mm_alltasks_mult_ks}. On the fast-judge score, nanoHEP-multi's outputs are within $0.05$ of chance on $779$ of the $1302$ directions, and its median is $0.54$. The most difficult tasks are those that generate the calorimeter-cell image, with a median of $0.58$ against $0.52$ for the rest. HEP4M-multi's outputs match truth at the level of marginal distribution - with median marginal Kolmogorov-Smirnov distance of $0.05$ against nanoHEP-multi's $0.02$ and multiplicities close to the truth. Given the event's inputs, however, its outputs are separable from the truth on almost every direction, with a median fast-judge score of $0.95$. That is, HEP4M-multi reproduces the marginals while missing event-level structure. nanoHEP-multi scores lower than HEP4M-multi on all $1302$ directions.

\subsection{Cross-task and unseen-task generalisation} \label{sec:generalization} \label{sec:mm_transfer}

\begin{figure}[htbp]
 \centering
 \includegraphics[width=0.85\linewidth]{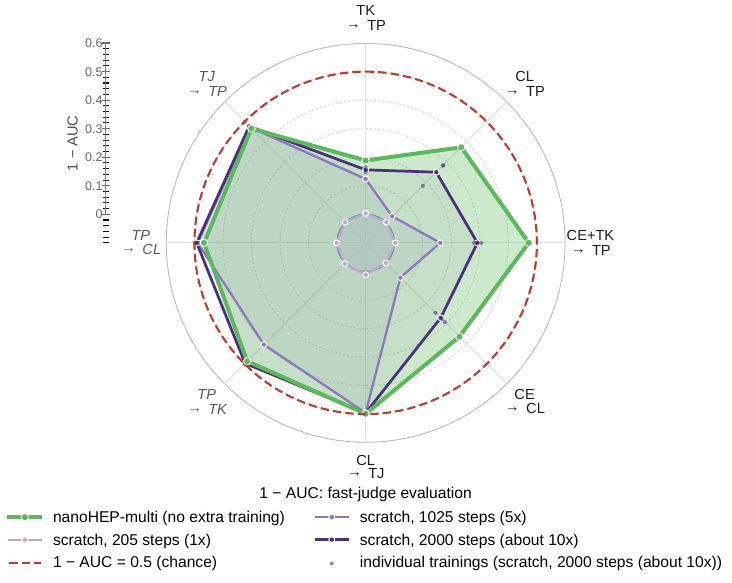}
 \caption{Typical tasks evaluated with nanoHEP-multi and with nanoHEP trained from scratch on each direction at nanoHEP-multi's on-task budget ($205$ steps), at five times ($1025$ steps) and at ten times that budget ($2000$ steps). The radius is $1 - \mathrm{AUC}$ of the fast-judge score of Sec.~\ref{sec:mm_alltasks}. Simulation-side directions are in italics and decoded by sampling at $T=1$; reconstruction-side directions are decoded by argmax. 
 TK = tracks, CL = topoclusters, CE = calorimeter cells, TP = truth particles, TJ = truth jets.
 }
 \label{fig:tail_transfer}
\end{figure}

Besides the oversampled particle flow direction, every task is drawn uniformly, at approximately one training sample in fifteen hundred. Is a direction seen this rarely learned at all, and does the rest of the matrix help it? We probe this first by comparing nanoHEP performance on eight tasks that were seen during training with equivalent models trained from scratch. The from-scratch model performance is averaged over three independent trainings with the same number of on-task steps that nanoHEP-multi received per task (Fig.~\ref{fig:tail_transfer}). At that budget nanoHEP-multi is less distinguishable from truth on all eight tasks as evaluated by the fast-judge score. At five times that budget nanoHEP-multi is still ahead on six of the eight directions, including all four reconstruction directions with a clear margin. At ten times that budget it remains ahead on the five reconstruction directions, and on the three generation directions, decoded by sampling, the from-scratch models are slightly closer to the truth. For the two cell-input directions the comparison is matched in steps but not in tokens, since a cell input is four times longer. We posit that this is a prototypical case of task transfer: training across N tasks with compute M benefitting performance more than N models, each trained with M/N compute. This is a promising result that a particle physics foundation model may be more than a convenient tool, but actually more computationally efficient.

Interestingly, the multi-task training also affords an improvement in particle reconstruction if alternative input modalities are considered. Besides the standard tracks + topoclusters $\rightarrow$ truth particles direction (which is heavily oversampled), the training also includes alternative tasks that produce the same target, i.e. truth particles: If the cell information is additionally added to the inputs, i.e. the tracks + topoclusters + cells $\rightarrow$ truth particles direction, the model outperforms the standard particle flow task (IQR $0.070$ against $0.078$) as the model can additionally exploit low-level correlations extracted from the cell data.

\begin{table}[htbp]
 \centering
 \footnotesize
 \setlength{\tabcolsep}{3pt}
 \begin{tabular}{lcccc}
  \toprule
  & & \multicolumn{2}{c}{jet $p_T$ response} & classifier \\
  \cmidrule(lr){3-4}
  inputs & trained & median & IQR [$95\%$ interval] & AUC \\
  \midrule
  tracks + topoclusters (benchmark)                    & yes & $0.990$ & $0.0781$ [$0.0775$, $0.0787$] & $0.703$ \\
  \midrule
  tracks + cells                                       & yes & $1.006$ & $0.0875$ [$0.0867$, $0.0882$] & - \\
  \midrule
  tracks + topoclusters + cells                        & yes & $0.999$ & $0.0704$ [$0.0698$, $0.0710$] & \textbf{0.650} \\
  tracks + topoclusters + HGPflow particles            & yes & $0.994$ & $0.0752$ [$0.0745$, $0.0758$] & $0.712$ \\
  tracks + HGPflow particles + cells                   & yes & $0.999$ & $0.0732$ [$0.0726$, $0.0738$] & $0.672$ \\
  topoclusters + HGPflow particles + cells             & yes & $1.001$ & \textbf{0.0690} [$0.0684$, $0.0696$] & $0.669$ \\
  \midrule
  tracks + topoclusters + HGPflow particles + cells    & no  & $0.999$ & \textbf{0.0688} [$0.0682$, $0.0694$] & $0.662$ \\
  \midrule
  tracks + topoclusters + cells + truth jets           & no  & $0.993$ & $0.0305$ [$0.0303$, $0.0308$] & - \\
  \bottomrule
 \end{tabular}
 \caption{Particle flow by nanoHEP-multi from different input combinations, with no further training, argmax decoding: jet $p_T$ response median and IQR with the $95\%$ bootstrap interval ($1000$ event-level resamples). The trained column marks whether the task was seen during training. The AUC is the set-transformer test of Sec.~\ref{sec:metrics} conditioned on the benchmark inputs, five retrainings (spread $\le 0.009$).}
 \label{tab:input_ladder}
\end{table}

The nanoHEP model additionally exhibits generalisation to tasks that were not present at all during the training. In Table~\ref{tab:input_ladder}, we report two tasks with four input modalities, which goes beyond the maximum of three modalities combinations seen during training. Among them is an additional ``particle flow-like'' direction with tracks,  topoclusters, HGPflow particles and cells as input. Compared to the tracks+topoclusters+cells performance, this additionally improves reconstruction performance (IQR $0.069$). We hypothesize that the model is able to use the HGPflow particles (which are produced from tracks and topoclusters) as a basis which is then refined using the additional sensor-level information. To our knowledge, this is the first demonstration in particle physics of a pre-trained model, with no further training, doing at least as well on an input combination absent from training as on every trained one, although such zero-shot composition of modalities is well documented in machine learning~\cite{Zhang2023UnseenModality, Girdhar2023ImageBind}.

\subsection{Limitations} \label{sec:mm_limits} While the model appears to be able to incorporate cell-level information usefully when presented at inputs, it currently struggles when they are requested as an output modality: nanoHEP-multi's cells generated from truth particles are fully separable from the truth ($0.998$). This is measured against the tokenised truth, so the generative model, and not the tokeniser, is the source of this separation. The tokeniser adds a second, separate loss: raw cells and the same cells after a tokeniser round trip are separable with an AUC of $0.96$ (\ref{app:cellroundtrip}). We leave both to future work.

\section{Conclusions} \label{sec:discussion}

This work shows that suitably high-fidelity tokenisation makes training across a wide variety of tasks straightforward. Training was stable for both HEP4M and nanoHEP, with little hyperparameter tuning (\ref{subsec:training_inference}). We did not compare against training without tokens. Although both architectures are proofs of concept, HEP4M-pflow is within $11\%$ of the state of the art in jet resolution, and nanoHEP-pflow has the best jet resolution and the lowest classifier AUC on the particle flow task (tracks and topoclusters to particles; Table~\ref{tab:pflow_main}). For particle reconstruction from other inputs, nanoHEP-multi does better still, reaching an IQR of $0.069$ and an AUC of $0.66$ from four inputs (Table~\ref{tab:input_ladder}).

This work shows that some iteration appears \emph{necessary} to capture the full multimodal posterior. 
For tokenised sequences, autoregression is the simplest limit of such iterative prediction. Using the classifier at the object level, 
we have shown that a reconstruction with realistic spectra and jets can still be distinguished from the truth particles. Indeed, HGPflow - which by every other metric delivers a highly realistic particle flow reconstruction - produces a predicted particle point cloud that is easily distinguished from ground truth once the classifier can condition on the event (AUC $0.90$, against $0.69$ for nanoHEP-pflow). The cost of such iterative approaches is captured by a Pareto frontier such as the one in Fig.~\ref{fig:headline_pareto}, and we posit that, going forward, foundation models in particle physics should be compared on Pareto frontiers of fidelity against computational cost rather than on outright performance. 

We have explored a wide variety of ideas and phenomena observed in large language models, to see if they transplant cleanly to a physics foundation model. We see nanoHEP-multi transfer to directions it has rarely seen, beating models trained from scratch at its own on-task budget, and matching them at ten times that budget (Sec.~\ref{sec:mm_transfer}). We see zero-shot generalisation to an unseen combination of four modalities, which matches or improves on every trained subset (Table~\ref{tab:input_ladder}). While limitations have been identified and need study in future work (Sec.~\ref{sec:mm_limits}), the broad capabilities of these multi-modal models even with sensor-level inputs, point toward a compelling future.

\section*{Acknowledgement}

\noindent M.K., J.K. and S.K. are supported by the US Department of Energy (DOE) under Grant No. DE-AC02-76SF00515, and by the American Science Cloud HEP Intelligent Data Activity TREASURE project within the US DOE Genesis Mission. L.H. is supported by BMFTR Project SciFM 05D25WO2 and by the European Research Council (ERC) under the European Union’s Horizon Europe research and innovation program grant agreement 101220713 (LEGO). D.M. was supported in this work by the Danish Data Science Academy, which is funded by the Novo Nordisk Foundation (NNF21SA0069429). B.H. was supported by the Excellence Cluster ORIGINS, funded by the Deutsche Forschungsgemeinschaft (DFG, German Research Foundation) under Germany's Excellence Strategy-EXC-2094-390783311.  NK and EG are supported by the Minerva Grant, the Knell Family Institute for Artificial Intelligence, The Benoziyo Center for High Energy Physics, a BSF Grant, and the Krenter-Perinot Center for High Energy Particle Physics.

\section*{Code and data availability}

\noindent The code is available at \url{https://github.com/FM-for-HEP/hep-any2any} under the Apache-2.0 licence. It contains both models, the tokenisers, training and inference for any choice of input and output modalities, and the evaluation metrics, including the classifier two-sample test. A short set of tests checks an installation against the release: the released checkpoints reproduce reference predictions on test events, a training loop runs on the released data, and the released tokenisers reproduce the released tokens.

\noindent The data are available on Zenodo at \url{https://doi.org/10.5281/zenodo.22917591} under a CC-BY-4.0 licence. The record holds the tokenised training, validation and test sets for all seven modalities ($88.9$M, $100$k and $100$k events), the seven tokeniser checkpoints, the raw COCOA files and HGPflow predictions for the validation and test sets, and the trained checkpoint of every model in Tables~\ref{tab:pflow_main} and~\ref{tab:sim_main}. A loader in the code repository rebuilds the exact token layout used for training.

\noindent

\bibliography{super_res} \bibliographystyle{unsrt}

\newpage
\newpage

\clearpage
\appendix \section{Additional task results} \label{app:tasks}

We run inference with nanoHEP-multi and HEP4M-multi on many more directions than the handful in the main text, and collect those evaluations here. Unless stated otherwise, the directions below were trained only through the multimodal sampler (Sec.~\ref{subsec:tasks}).

\subsection{Reconstruction refinement: HGPflow particles $\rightarrow$ truth particles} \label{app:unfold} 
\begin{figure}[htbp]
 \centering
 \includegraphics[width=\linewidth]{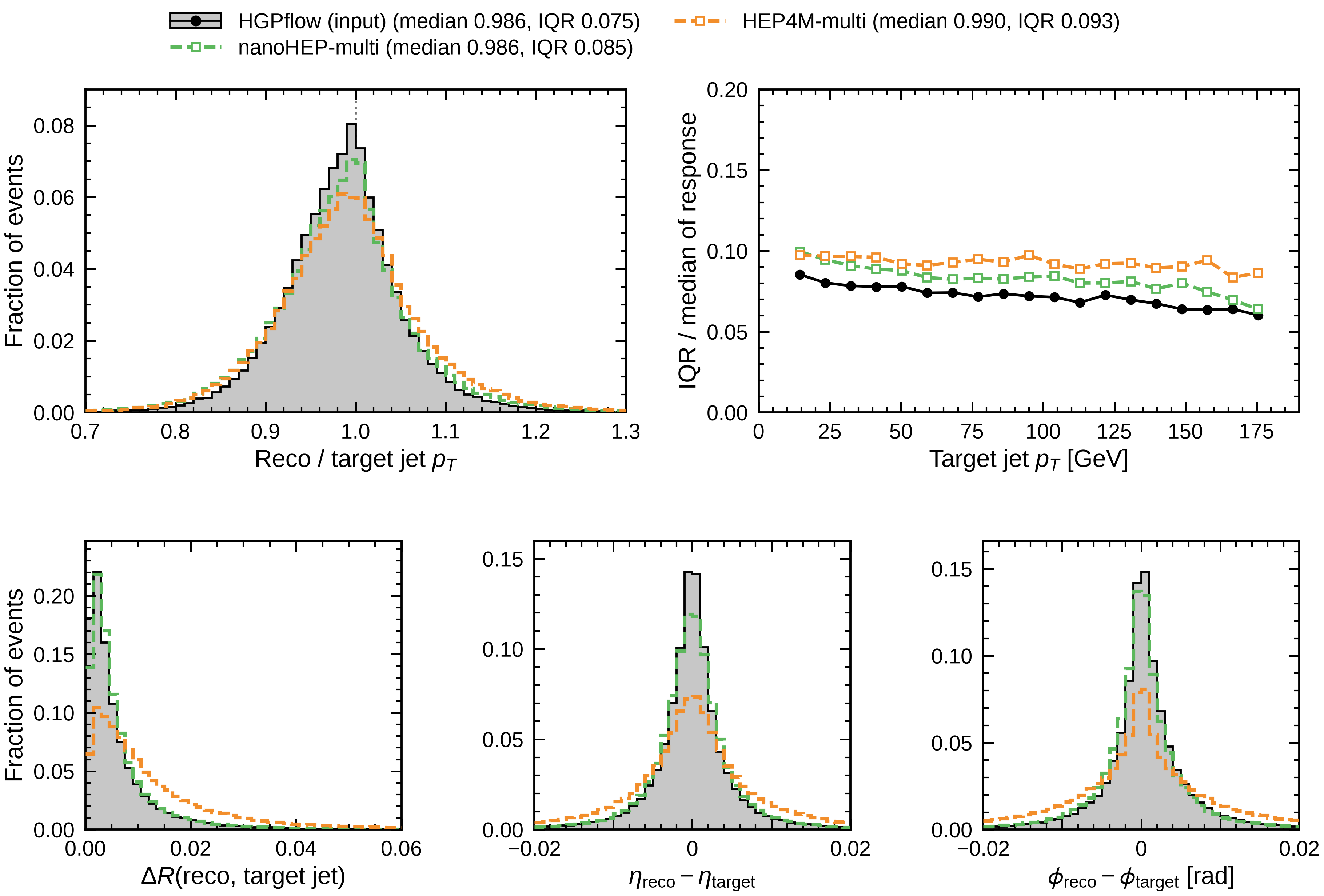}
 \caption{Jet $p_T$ response, response IQR/median versus target jet $p_T$, and the $\Delta R$, $\eta$ and $\phi$ residuals between reconstructed and target jets for the reconstruction-refinement task.}
 \label{fig:mm_unfold}
\end{figure}

Here the input is HGPflow's particles and the output the truth particles of the same event, so the task is to refine an existing reconstruction (Fig.~\ref{fig:mm_unfold}). The jet $p_T$ response IQR is $0.085$ for nanoHEP-multi and $0.093$ for HEP4M-multi. Neither is an improvement on the resolution of the HGPflow input.

\subsection{Flash simulation: truth particles $\rightarrow$ HGPflow particles} \label{app:flashsim} 
\begin{figure}[htbp]
 \centering
 \includegraphics[width=\linewidth]{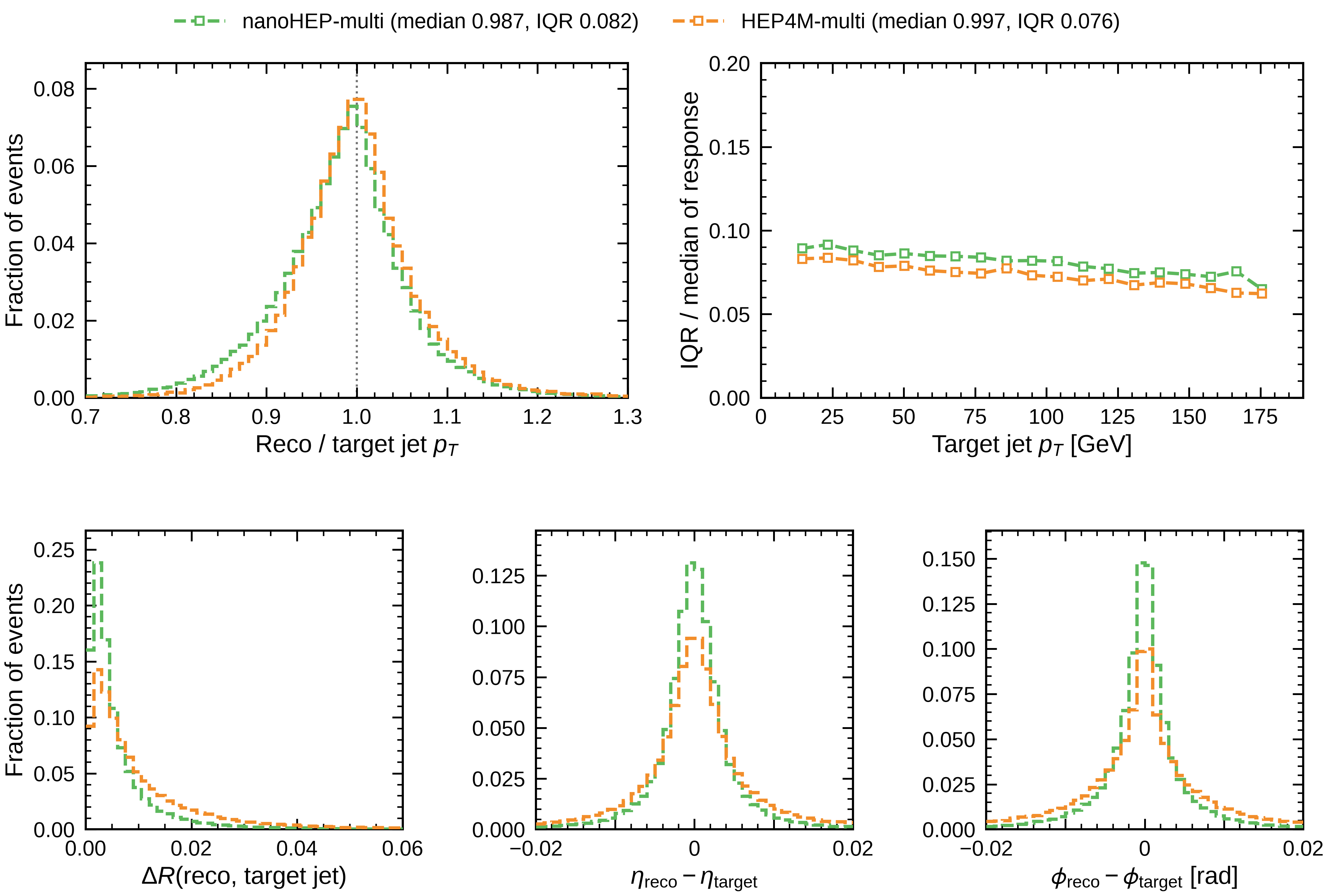}
 \caption{Flash simulation (truth particles $\rightarrow$ HGPflow particles), argmax decoding: jet $p_T$ response, response IQR/median versus target jet $p_T$, and the $\Delta R$, $\eta$ and $\phi$ residuals between reconstructed and target jets.}
 \label{fig:mm_flashsim}
\end{figure}

We run inference from truth particles directly to HGPflow particles, skipping the detector, as in a fast simulation. Here HEP4M-multi has the better jet response, with an IQR of $0.076$ against $0.082$ for nanoHEP-multi (Fig.~\ref{fig:mm_flashsim}). This is the one direction in this paper where the parallel model is ahead.

\subsection{Jet simulation: truth jet $\rightarrow$ truth particles} \label{app:jetsim} 
\begin{figure}[htbp]
 \centering
 \includegraphics[width=\linewidth]{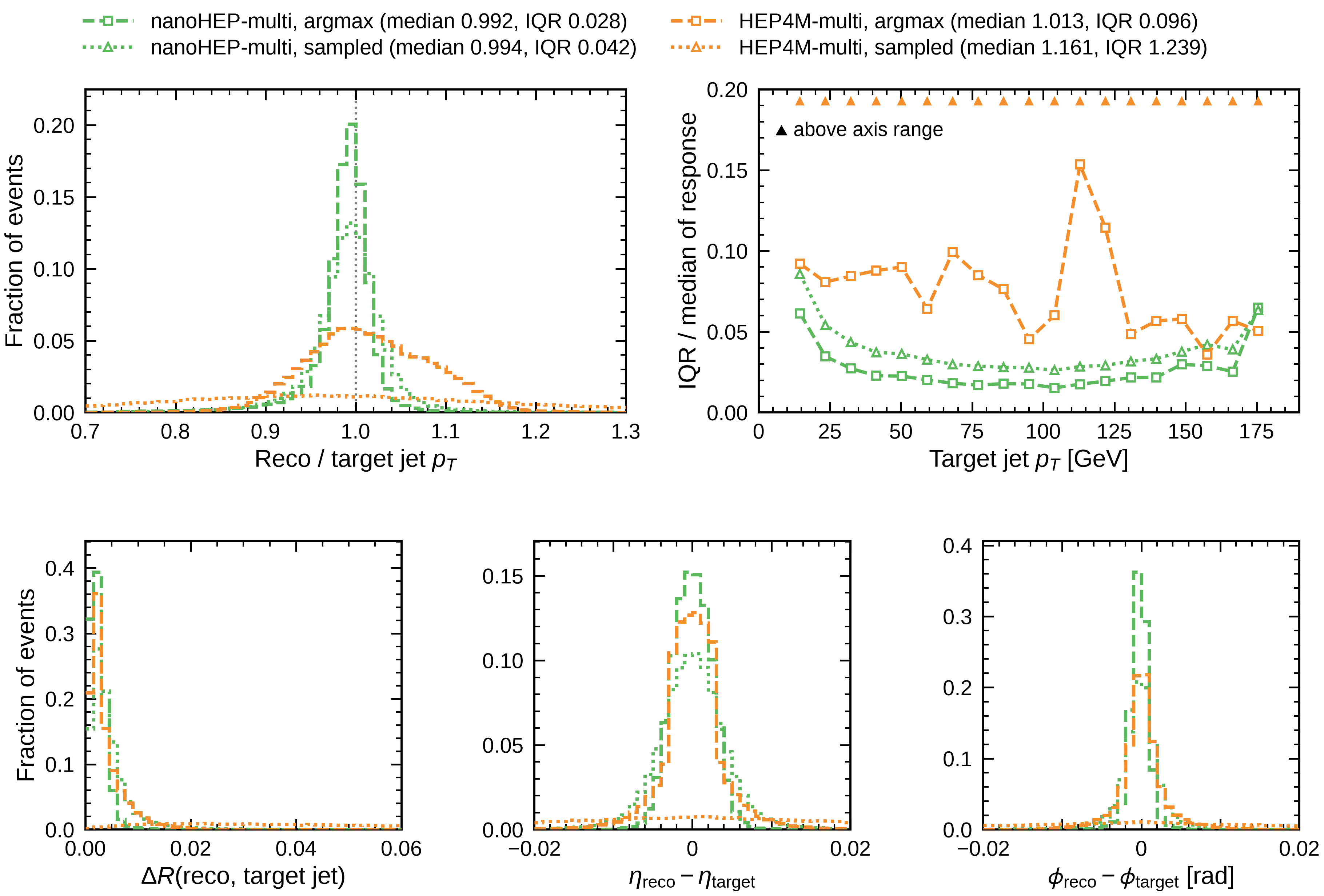}
 \caption{Jet $p_T$ response, response IQR/median versus target jet $p_T$, and the $\Delta R$, $\eta$ and $\phi$ residuals between reconstructed and target jets for the jet-simulation task, decoded with argmax and with temperature-1 sampling.}
 \label{fig:mm_jetsim}
\end{figure}

In this direction a single input has many valid outputs, so we compare argmax with sampling at temperature $1$. We generate a jet's constituent particles from the jet alone (Fig.~\ref{fig:mm_jetsim}). With argmax, the IQR is $0.028$ for nanoHEP-multi and $0.096$ for HEP4M-multi. With sampling, nanoHEP-multi's IQR is $0.042$, while HEP4M-multi's response is broad (IQR $1.24$) and lies above the plotted range.

\subsection{Charged-energy subtraction: cell $\rightarrow$ charged fraction} \label{app:chesub} 
\begin{figure}[htbp]
 \centering
 \includegraphics[width=\linewidth]{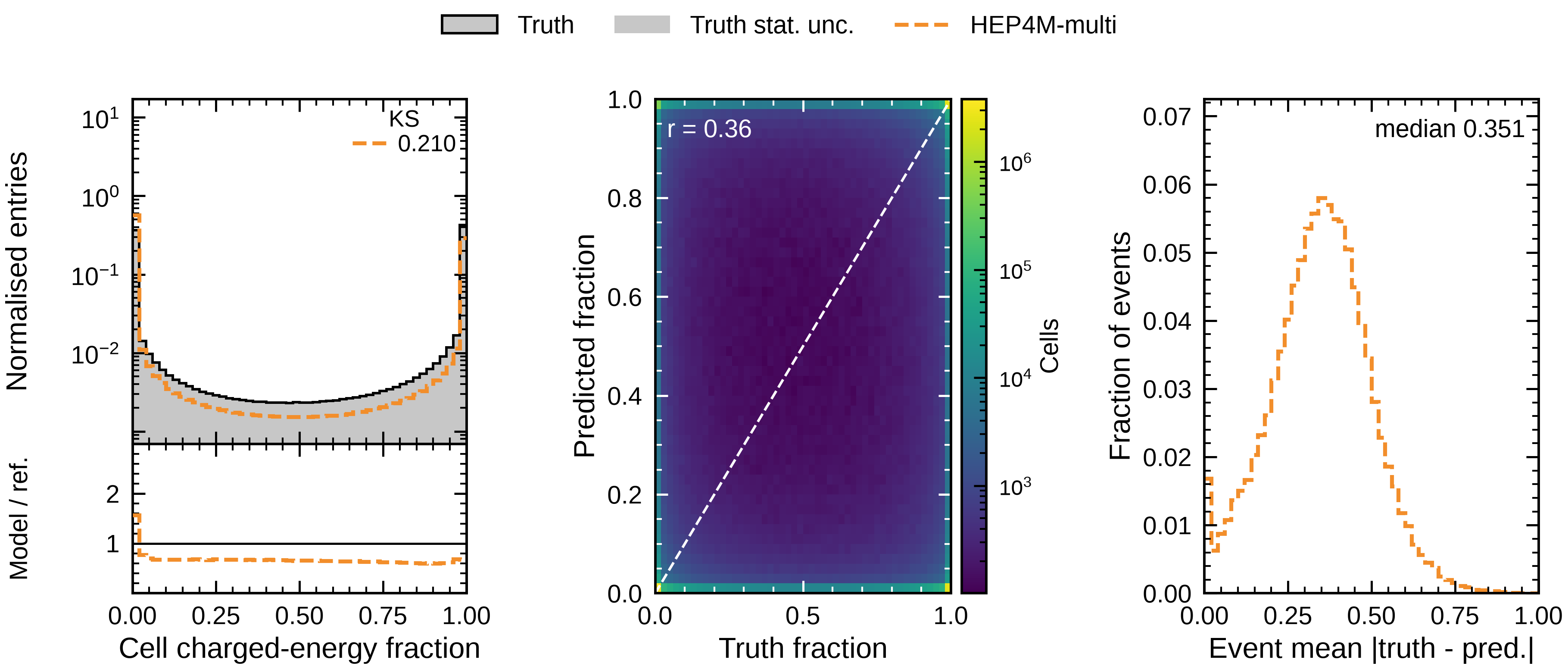}
 \caption{Charged-energy subtraction by HEP4M-multi: the charged-fraction distribution for truth and prediction, the per-cell truth versus predicted fraction, and the per-event mean absolute error.}
 \label{fig:mm_chesub}
\end{figure}

We run inference with HEP4M-multi to predict each calorimeter cell's charged-energy fraction from the raw cell image (Fig.~\ref{fig:mm_chesub}). The truth fraction is strongly bimodal: $35\%$ of cells are below $0.01$ and $42\%$ above $0.99$. We therefore compare against two baselines, the best constant and the truth fraction of a randomly chosen other cell, with mean absolute errors per cell of $0.45$ and $0.49$. For HEP4M-multi the error is $0.35$, with a per-cell correlation to the truth of $0.36$. Of course, this is weak next to the reconstruction directions of Sec.~\ref{sec:headline}. We include it to show that the task is partly learned, not that it is solved.

\subsection{Detector simulation: additional detail} \label{app:simdetail} 
\begin{figure}[htbp]
 \centering
 \includegraphics[width=\linewidth]{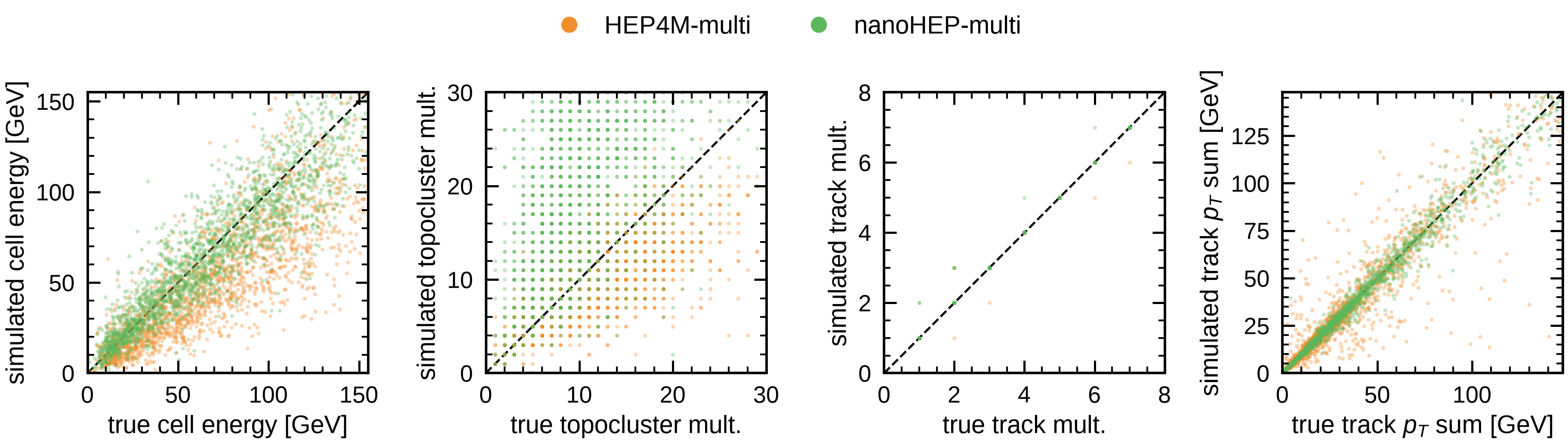}
 \caption{Per-event aggregates of the simulated detector response versus truth (sampled decoding): total cell energy, topocluster multiplicity, track multiplicity, and summed track $p_T$, for nanoHEP-multi and HEP4M-multi.}
 \label{fig:mm_sim_scatters}
\end{figure}

\begin{figure}[htbp]
 \centering
 \includegraphics[width=0.49\linewidth]{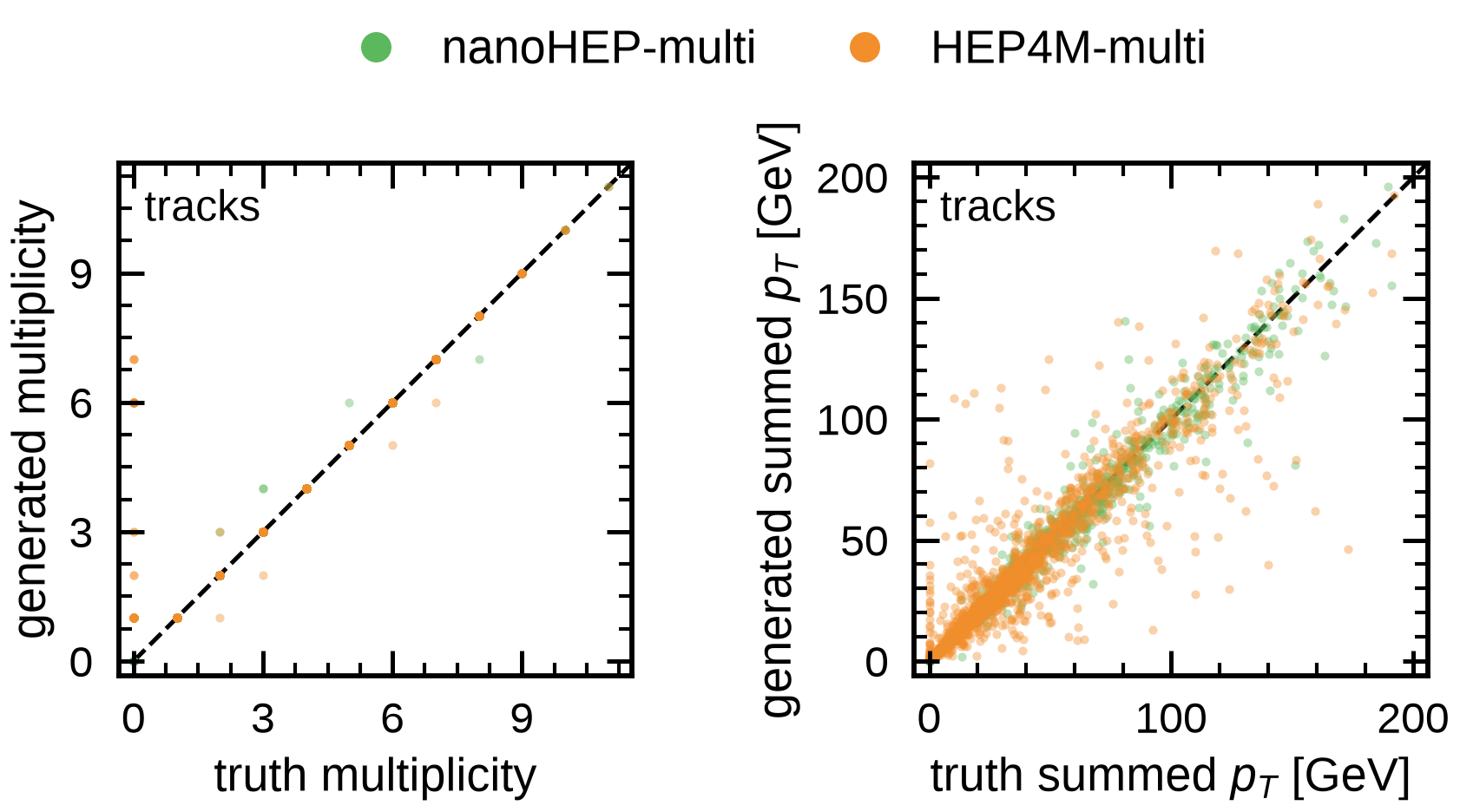}
 \includegraphics[width=0.49\linewidth]{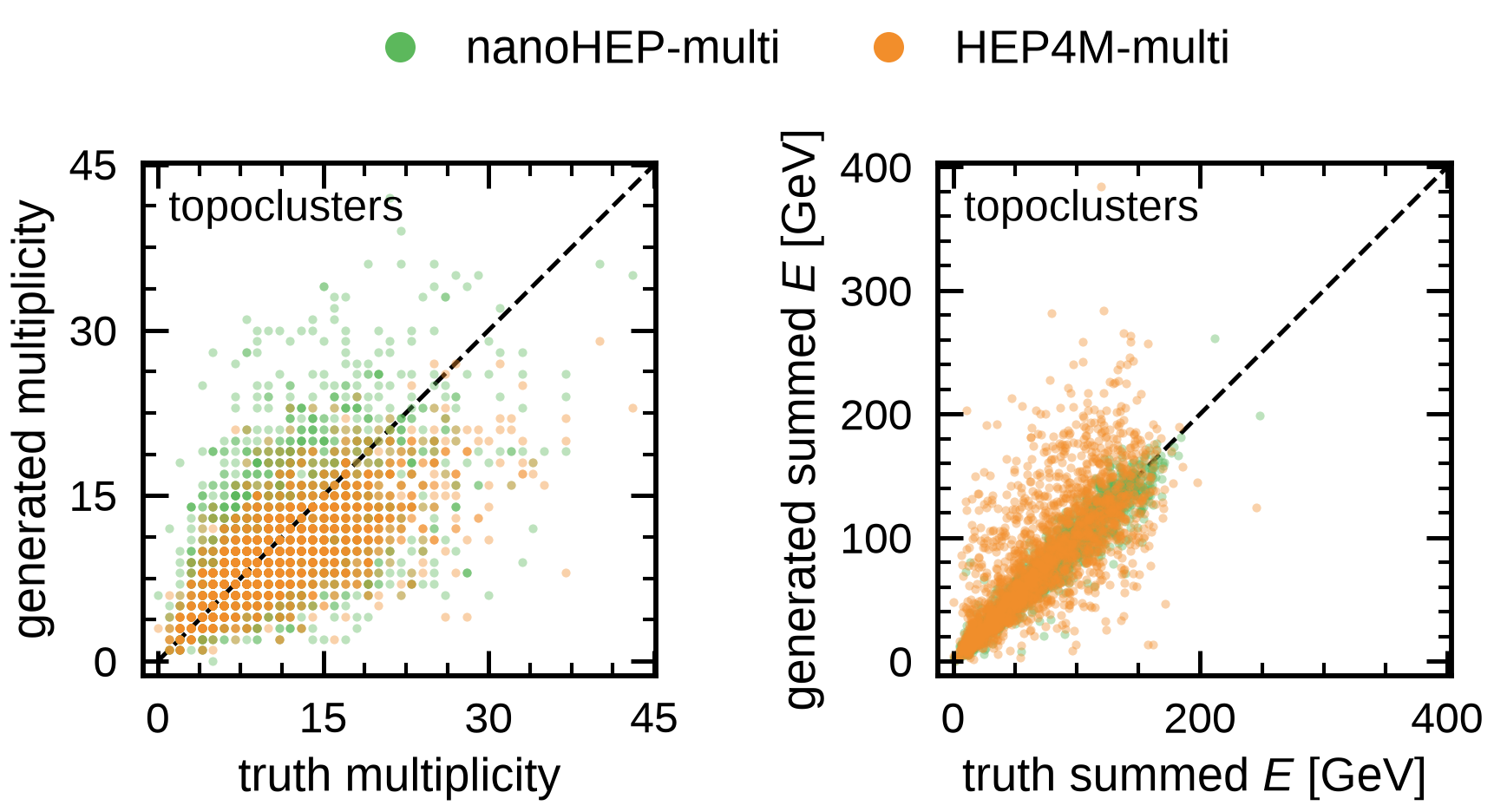}
 \caption{Per-event truth versus generated multiplicity and summed magnitude for particles $\rightarrow$ tracks (left) and particles $\rightarrow$ topoclusters (right).}
 \label{fig:decomp_scatter}
\end{figure}

\begin{figure}[htbp]
 \centering
 \includegraphics[width=\linewidth]{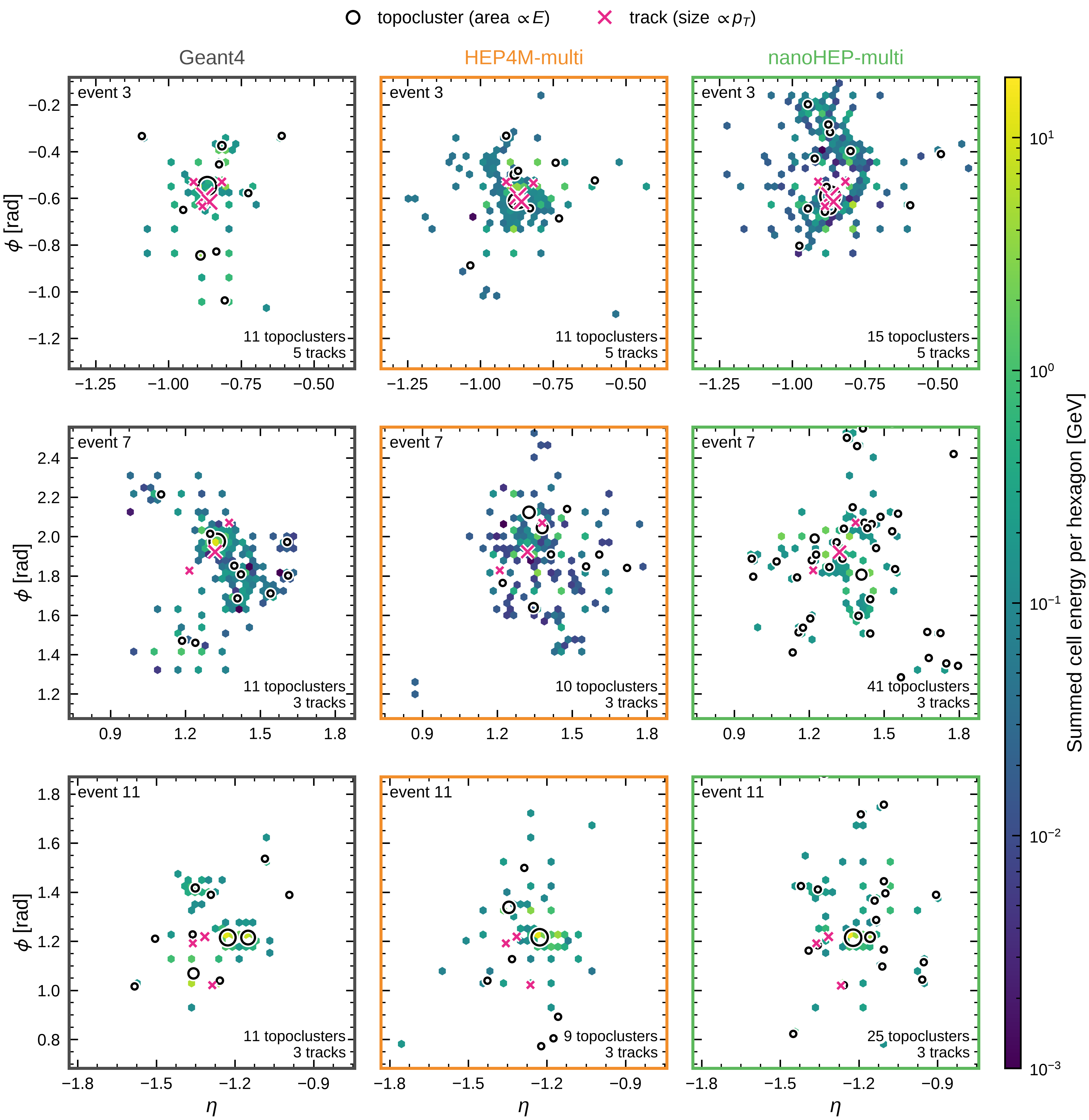}
 \caption{Example events in $\eta$-$\phi$: truth, HEP4M-multi and nanoHEP-multi, sampled decoding. Cell energies are shown as heatmaps, topoclusters as circles (area proportional to energy) and tracks as crosses (size proportional to $p_T$).}
 \label{fig:mm_sim_displays}
\end{figure}

For detector simulation (particles $\rightarrow$ cells, topoclusters and tracks), the per-event aggregates are in Fig.~\ref{fig:mm_sim_scatters}, the per-modality aggregates in Fig.~\ref{fig:decomp_scatter} and example events in Fig.~\ref{fig:mm_sim_displays}. The track and topocluster marginals are in Sec.~\ref{sec:generative} (Figs.~\ref{fig:gen_track} and~\ref{fig:gen_topo}). In the three-output direction, nanoHEP-multi's topocluster multiplicity is about twice the truth (median per-event ratio $2.0$), while its track multiplicity matches.

\subsection{Topoclustering} \label{app:topoclus} 
\begin{figure}[htbp]
 \centering
 \includegraphics[width=\linewidth]{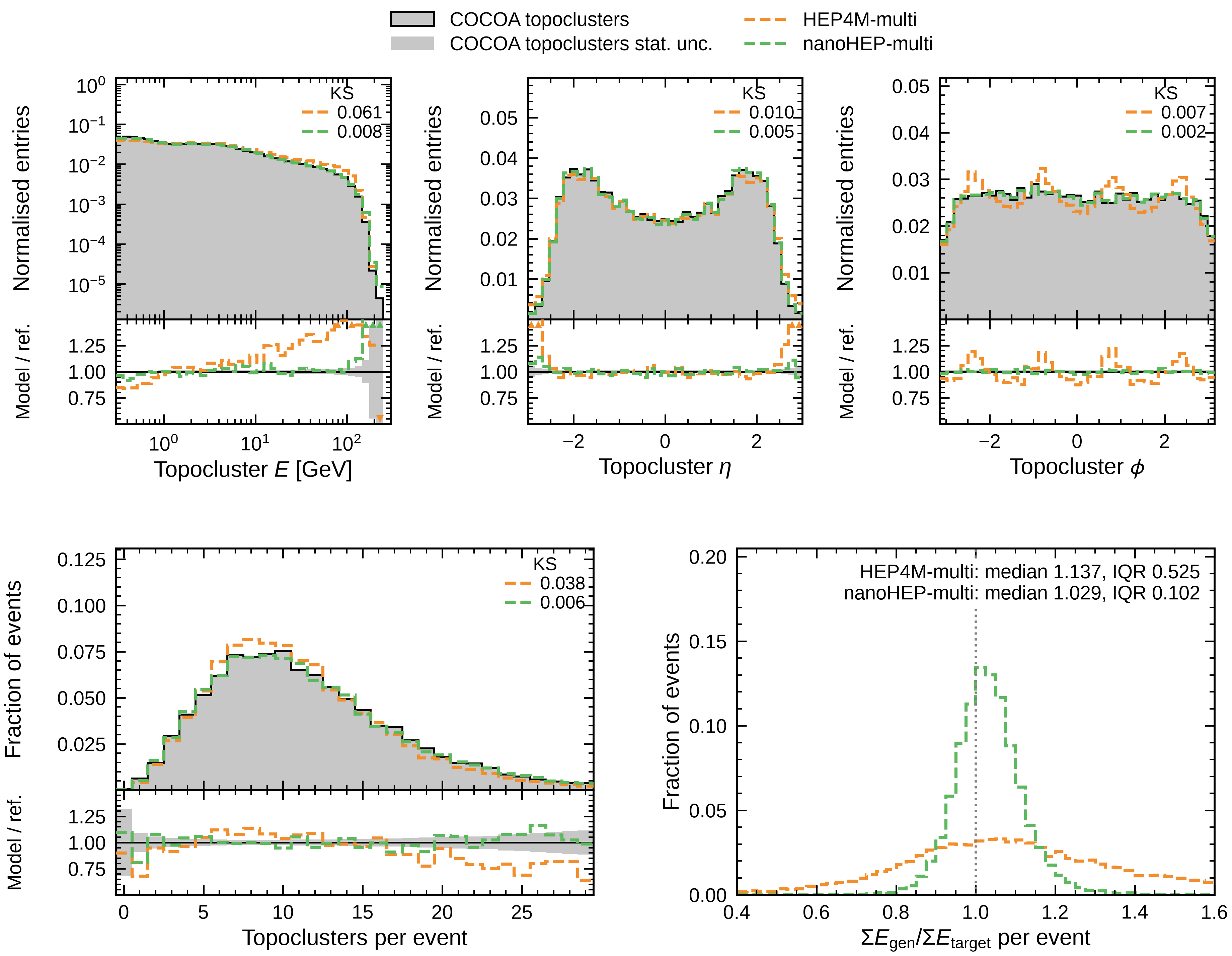}
 \caption{Topoclustering (tracks and cells $\rightarrow$ topoclusters) by nanoHEP-multi and HEP4M-multi: single-object marginals, topocluster multiplicity, and the per-event summed-energy response.}
 \label{fig:topoclus_oob}
\end{figure}

We run topoclustering (tracks and cells $\rightarrow$ topoclusters) with nanoHEP-multi and HEP4M-multi (Fig.~\ref{fig:topoclus_oob}). With sampled decoding, nanoHEP-multi's summed-energy response against the COCOA topoclusters has a median of $1.03$ and an IQR of $0.10$, and its cluster count is within $1\%$ of the target.

\subsection{Calorimeter-cell tokeniser round trip} \label{app:cellroundtrip} 
\begin{figure}[htbp]
 \centering
 \includegraphics[width=\linewidth]{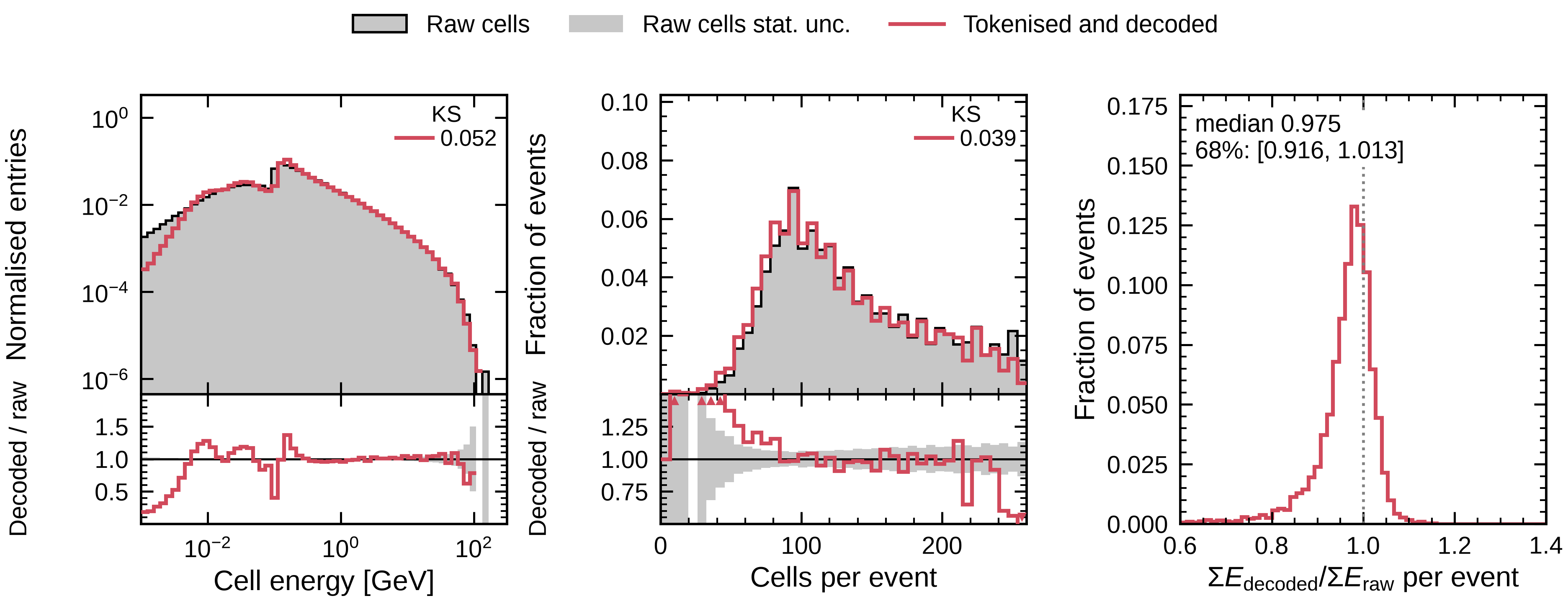}
 \caption{Calorimeter cells before and after a tokeniser round trip (tokenised and decoded back, no model involved): cell energy spectrum (left), cells per event (middle), and the ratio of decoded to raw summed cell energy per event (right).}
 \label{fig:cell_roundtrip}
\end{figure}

Comparisons against the tokenised truth hide the loss in the tokeniser itself. In Fig.~\ref{fig:cell_roundtrip} we compare raw calorimeter cells with the same cells tokenised and decoded back, with no model in between. Below about $5$~MeV the decoded spectrum is too low, so an event has about $4\%$ fewer cells after the round trip. The summed energy per event is $2.5\%$ low at the median, and a sixth of events are more than $8\%$ low. Raw and round-tripped cells are separable by a set-transformer classifier (Sec.~\ref{sec:metrics}) given the cells alone, with an AUC of $0.96$.

\subsection{All directions: multiplicity and marginals} \label{app:alltasks_mult_ks} 
\begin{figure}[htbp]
 \centering
 \includegraphics[width=\linewidth]{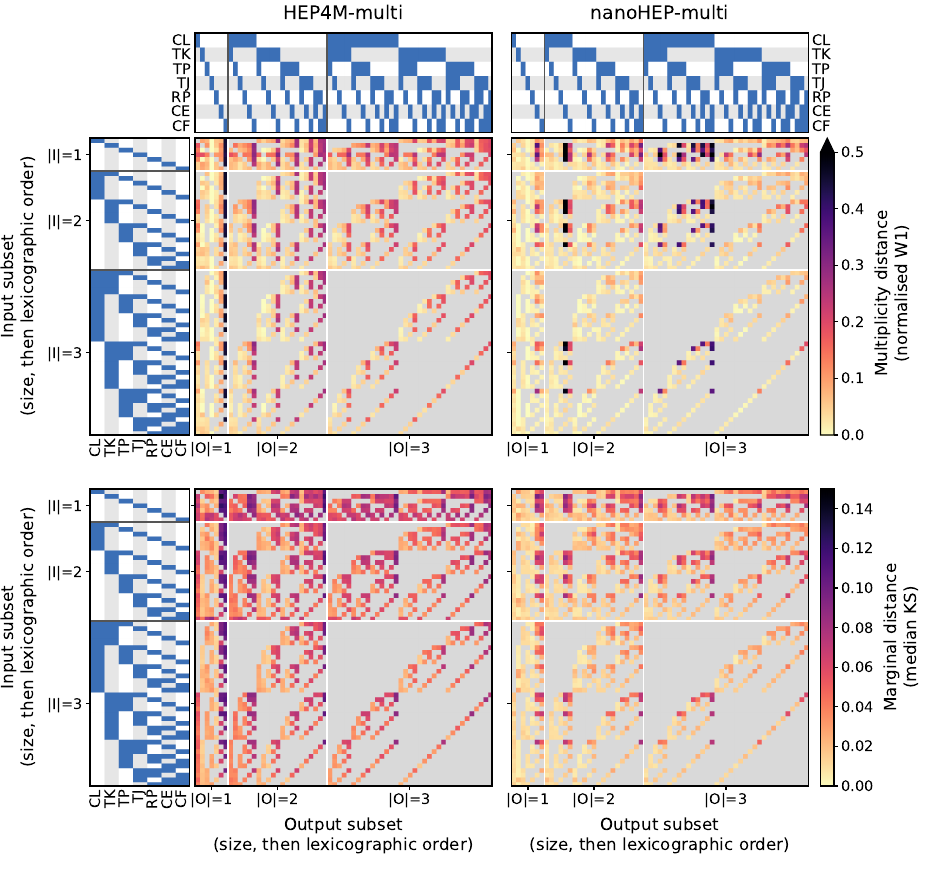}
 \caption{All $1302$ task directions for HEP4M-multi (left) and nanoHEP-multi (right): normalised multiplicity Wasserstein-1 (top) and median marginal Kolmogorov-Smirnov distance (bottom), on the grid of Fig.~\ref{fig:mm_alltasks}.}
 \label{fig:mm_alltasks_mult_ks}
\end{figure}

For the $1302$ directions of Sec.~\ref{sec:mm_alltasks}, the two measures not shown in Fig.~\ref{fig:mm_alltasks} are in Fig.~\ref{fig:mm_alltasks_mult_ks}: the normalised multiplicity Wasserstein-1 distance and the median marginal Kolmogorov-Smirnov distance.

\subsection{Additional particle flow and simulation views} 
\begin{figure}[htbp]
 \centering
 \includegraphics[width=\linewidth]{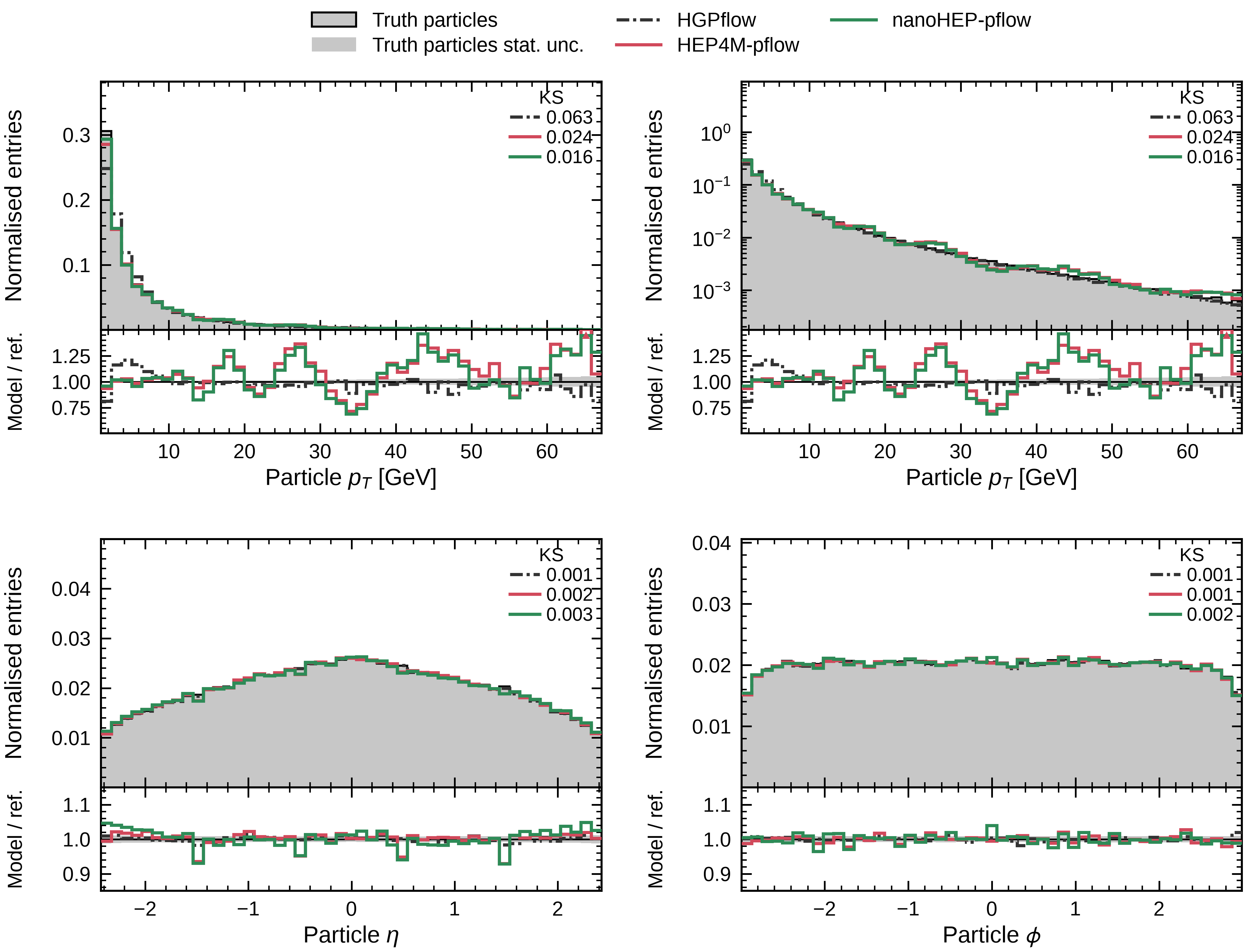}
 \caption{Per-particle $p_T$, $\eta$, and $\phi$ distributions for HGPflow, HEP4M-pflow, and nanoHEP-pflow.}
 \label{fig:headline_marginals}
\end{figure}

\begin{figure}[htbp]
 \centering
 \includegraphics[width=0.7\linewidth]{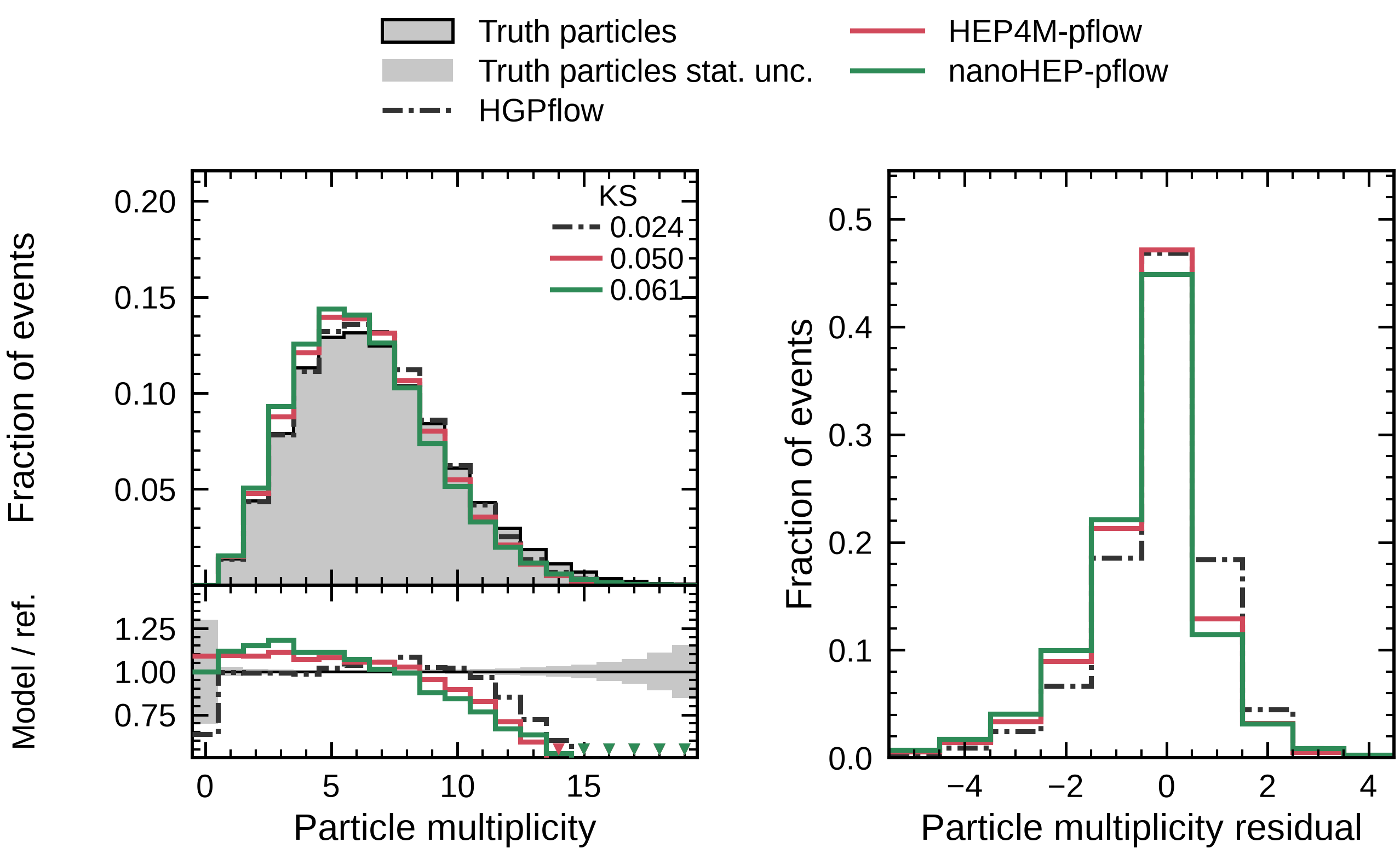}
 \caption{Per-event particle multiplicity for HGPflow, HEP4M-pflow, and nanoHEP-pflow.}
 \label{fig:headline_cardinality}
\end{figure}

\begin{figure}[htbp]
 \centering
 \includegraphics[width=\linewidth]{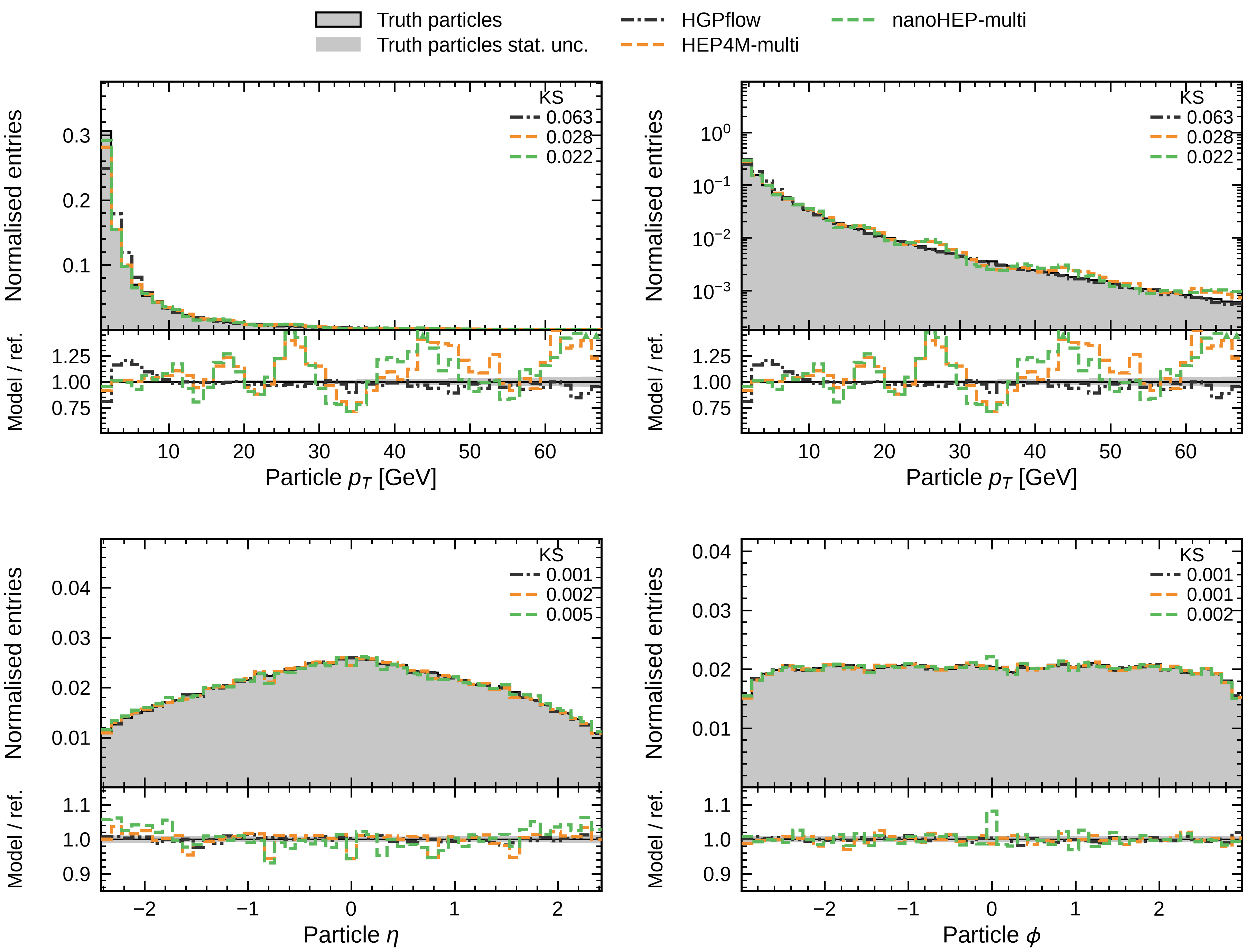}
 \caption{Per-particle $p_T$, $\eta$, and $\phi$ distributions for HGPflow, HEP4M-multi and nanoHEP-multi on particle flow.}
 \label{fig:mm_pflow_marg}
\end{figure}

\begin{figure}[htbp]
 \centering
 \includegraphics[width=\linewidth]{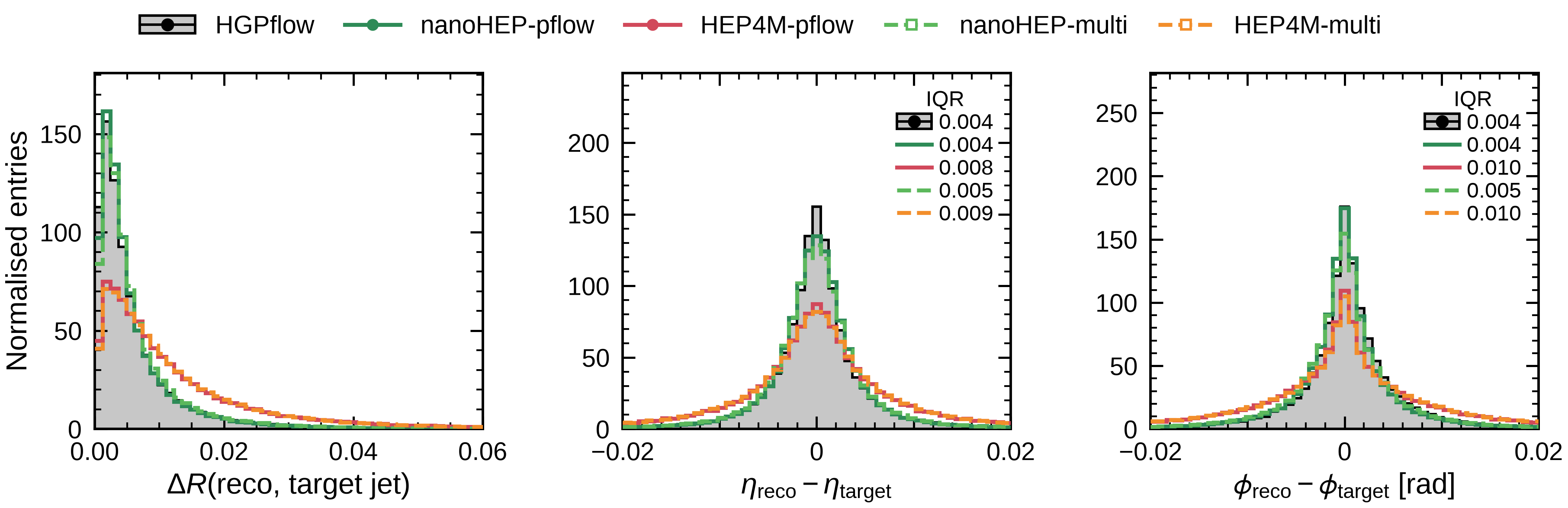}
 \caption{$\Delta R$, $\Delta\eta$ and $\Delta\phi$ between the reconstructed and target jets on particle flow, for HGPflow and the four particle flow models of Table~\ref{tab:pflow_main}.}
 \label{fig:headline_angles}
\end{figure}

\begin{figure}[htbp]
 \centering
 \includegraphics[width=0.49\linewidth]{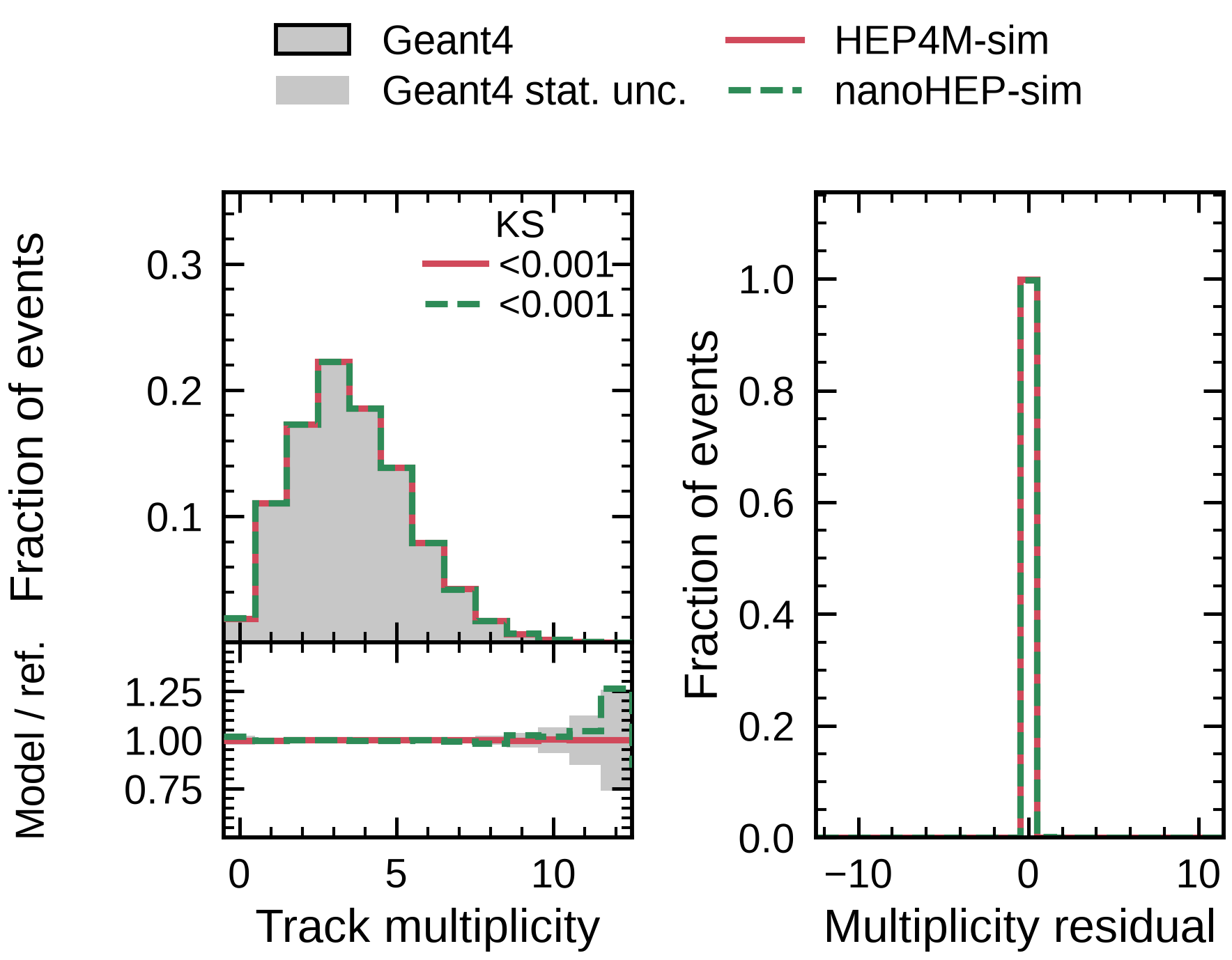}
 \includegraphics[width=0.49\linewidth]{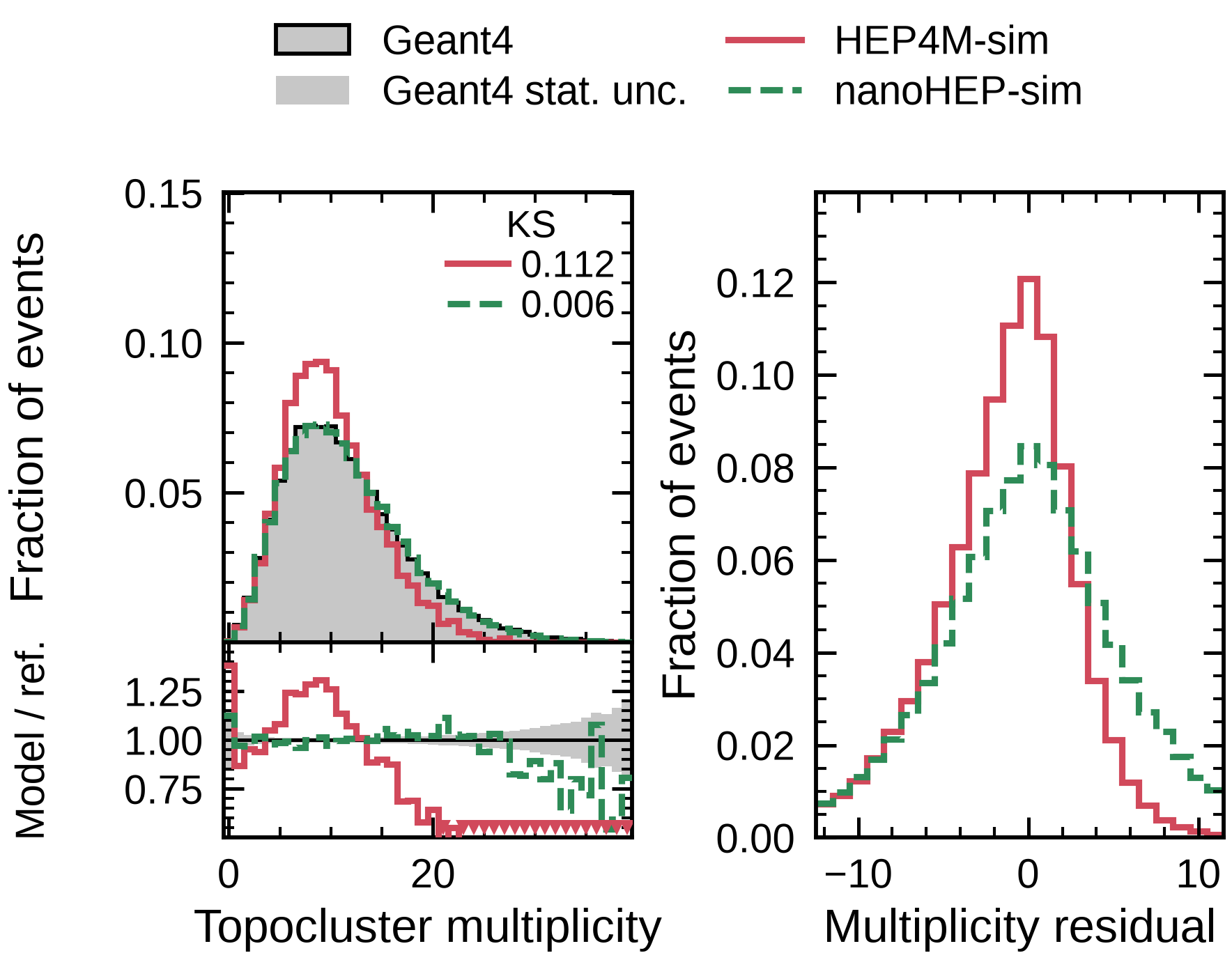}
 \caption{Per-event track (left) and topocluster (right) multiplicity, target versus generated (sampled decoding).}
 \label{fig:gen_card}
\end{figure}

For the single-task particle flow models of Sec.~\ref{sec:headline}, the per-particle marginals are in Fig.~\ref{fig:headline_marginals} and the particle multiplicity in Fig.~\ref{fig:headline_cardinality}. The same marginals for the multi-task models are in Fig.~\ref{fig:mm_pflow_marg}. The angular distances between reconstructed and target jets for all five particle flow models are in Fig.~\ref{fig:headline_angles}, and the track and topocluster multiplicities of the simulation task (Sec.~\ref{sec:generative}) in Fig.~\ref{fig:gen_card}.

 \section{Decoding studies: argmax versus sampling}
\label{app:decoding}

\begin{figure}[htbp]
 \centering
 \includegraphics[width=\linewidth]{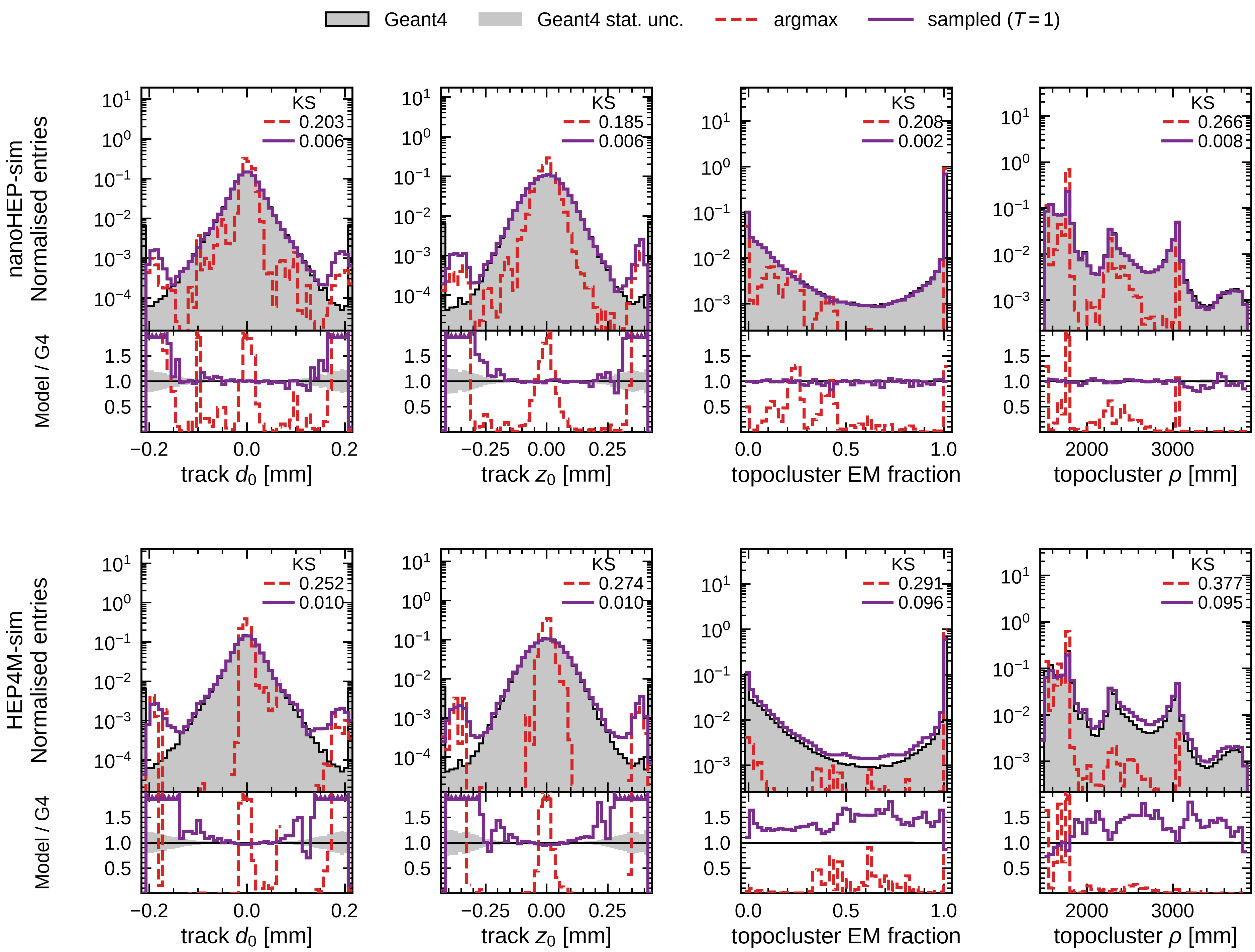}
 \caption{Track $d_0$ and $z_0$ and topocluster EM fraction and barycentre radius $\rho$ in detector simulation, for nanoHEP-sim (top) and HEP4M-sim (bottom), decoded by argmax and by sampling at $T=1$, against Geant4. The axes span the tokeniser's range; the $1.3\%$ of Geant4 tracks outside it are in the edge bins.}
 \label{fig:argmax_vs_sampled}
\end{figure}

\begin{figure}[htbp]
 \centering
 \includegraphics[width=\linewidth]{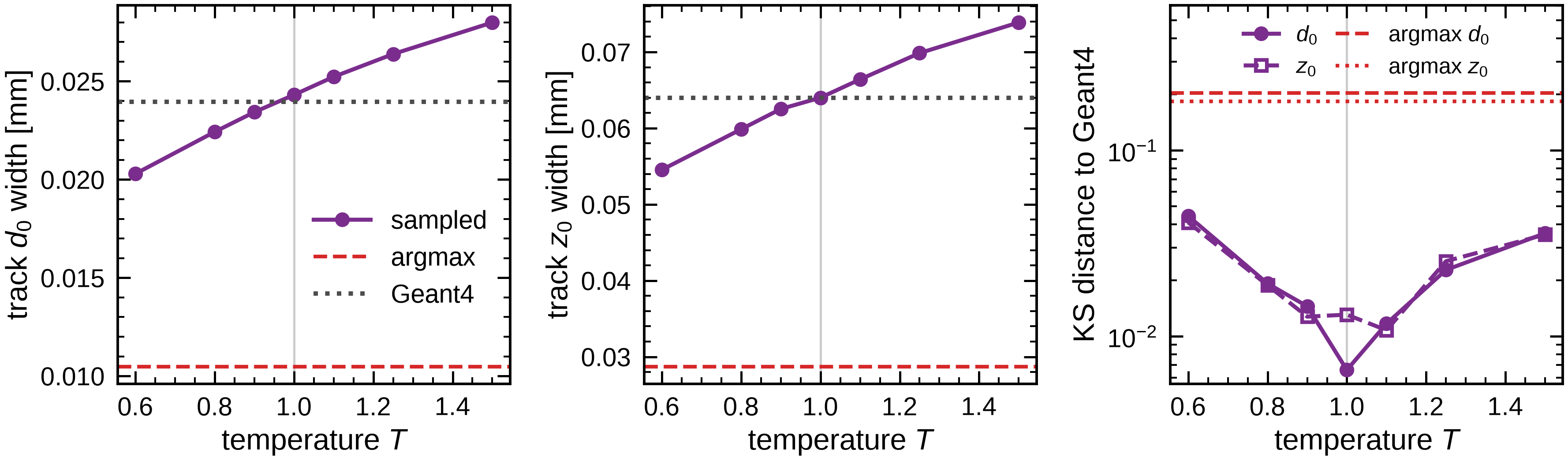}
 \caption{Track $d_0$ and $z_0$ widths (half the 16 to 84\% interval) from nanoHEP-sim, and their Kolmogorov-Smirnov distance to Geant4, versus sampling temperature $T$, compared with argmax decoding and with Geant4.}
 \label{fig:temp_scan}
\end{figure}

With argmax decoding, the sharply peaked marginals collapse for both models: the track impact parameters $d_0$ and $z_0$ are far too narrow (Fig.~\ref{fig:argmax_vs_sampled}). Sampling at temperature $T=1$ mitigates the collapsed widths for both models. At $T=1$ the tokens are sampled from the predictive distribution fit in training by maximum likelihood. In a scan of the temperature (Fig.~\ref{fig:temp_scan}), the widths are a smooth function of $T$ and match the Geant4 widths to within about $1\%$ at $T=1$.

Beyond single-feature widths, the two models are different. In nanoHEP each token is sampled conditioned on all previous tokens, so learned constraints are kept under sampling. The sampled $(\cos\phi, \sin\phi)$ pairs of nanoHEP-sim satisfy $\cos^2\phi + \sin^2\phi = 1.000 \pm 0.011$ for tracks and $1.000 \pm 0.006$ for topoclusters. In HEP4M every token is sampled independently given the input (Eq.~\ref{eq:parallel_factorisation}), so the identity is not enforced. Before the renormalisation of Sec.~\ref{sec:metrics}, the sum is $0.97 \pm 0.21$ for tracks and $0.97 \pm 0.30$ for topoclusters. HEP4M-sim's sampled topocluster marginals are also too flat. Between the two peaks of the EM fraction, and between the peaks of $\rho$, the density is about $1.4$ times that of Geant4 (Fig.~\ref{fig:argmax_vs_sampled}, bottom row, right two panels). The single-object marginals under sampled decoding are in Figs.~\ref{fig:gen_track} and \ref{fig:gen_topo}.

\begin{figure}[htbp]
 \centering
 \includegraphics[width=\linewidth]{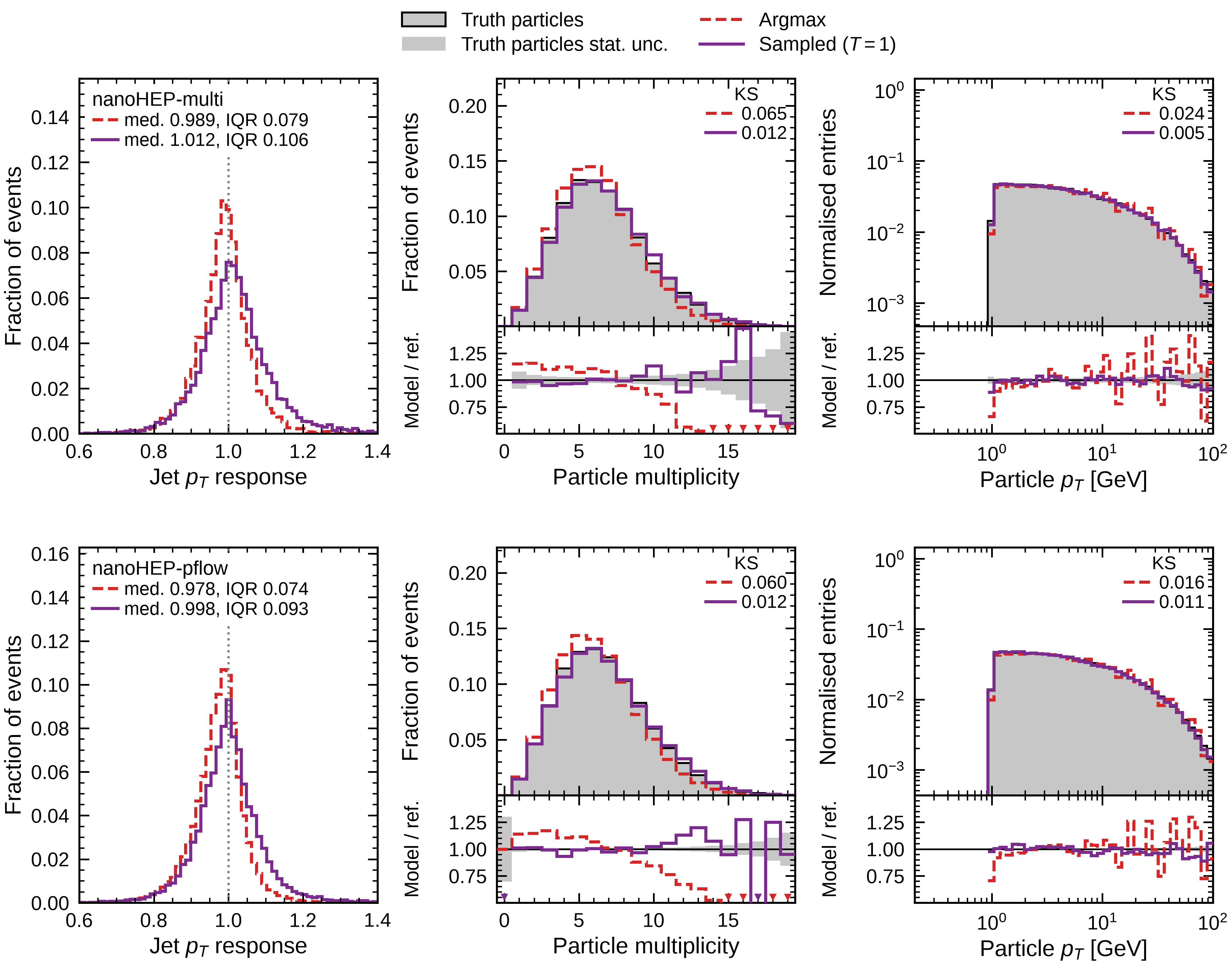}
 \caption{Particle flow with argmax and temperature-1 sampled decoding, for nanoHEP-multi (top row) and nanoHEP-pflow (bottom row): jet $p_T$ response, particle multiplicity, and particle $p_T$ spectrum.}
 \label{fig:pflow_decode_mode}
\end{figure}

The previous comparison demonstrates the value of argmax for reconstruction and of a temperature of 1 for generation tasks. For nanoHEP-pflow on particle flow (Fig.~\ref{fig:pflow_decode_mode}), the jet response is tighter with argmax (IQR $0.074$, against the tokenised truth), but the median is biased ($0.978$) and the particle multiplicity collapsed. With sampling, the median is $0.998$ and the IQR wider ($0.093$). For nanoHEP-multi, the AUC is $0.52 \pm 0.01$ with sampling against $0.70$ with argmax, the same level as its sampled detector simulation. For the input combinations of Table~\ref{tab:input_ladder}, the AUC is $0.65$ to $0.71$ with argmax and about $0.55$ with sampling, at a cost of about $0.025$ in IQR. On particle flow, the resolution is better with argmax and the distributional fidelity better with sampling.

 \section{Two-pass HEP4M decoding} \label{app:conditioning} \label{subsec:cond_twopass}

In HEP4M every particle is sampled independently given the event's inputs (Eq.~\ref{eq:parallel_factorisation}). With a second decoder pass, the remaining particles can be conditioned on one of them. We test this on particle flow with a $64$M-parameter HEP4M trained on particle flow alone, not the $114$M HEP4M-pflow of Sec.~\ref{sec:headline}. We train two versions of it that differ only in the objective. The first is trained as usual. For the second, on $80\%$ of training events the first output slot holds the tokens of the true leading-$p_T$ particle. On the rest it is empty, so the same weights also serve unconditional inference.

In pass one we decode the event with the first model and keep its leading particle. In pass two we decode the event again with the second model, with that predicted particle in the first slot. No truth is used at inference, and all sampling is argmax.

\begin{figure}[htbp]
 \centering
 \includegraphics[width=\linewidth]{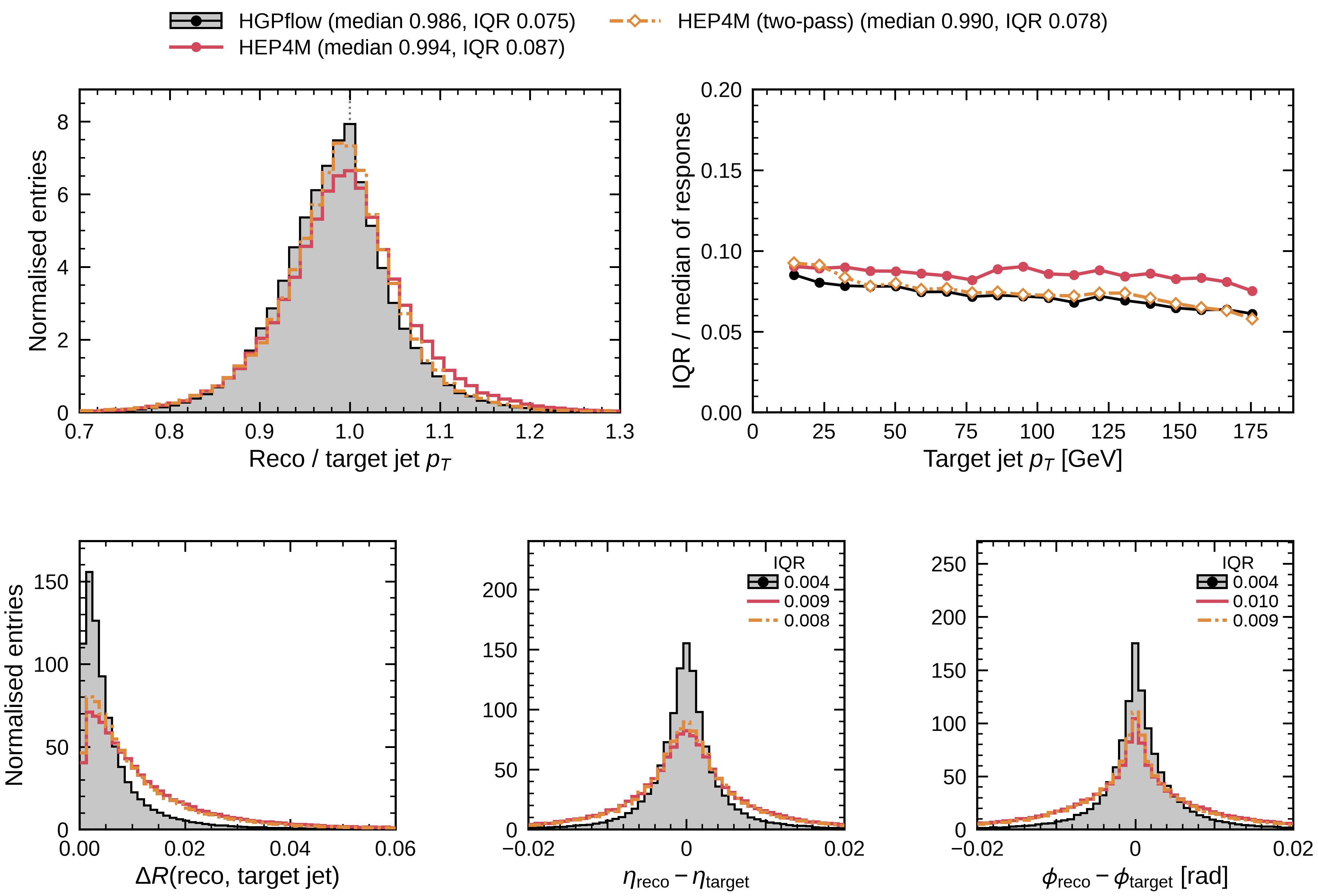}
 \caption{Jet $p_T$ response, response IQR/median versus target jet $p_T$, and $\Delta R$, $\Delta\eta$, $\Delta\phi$ for the single-pass and two-pass $64$M HEP4M, and HGPflow.}
 \label{fig:cond_jets}
\end{figure}

With two passes the jet $p_T$ response IQR is $0.0785$, against $0.0865$ for a single pass of the second model in its unconditional mode (Fig.~\ref{fig:cond_jets}). With HGPflow at $0.075$, this recovers about $70\%$ of the difference, at the cost of a second decoder pass (Fig.~\ref{fig:headline_pareto}).

 \section{Data and model scaling} \label{app:scaling}

\begin{figure}[htbp]
 \centering
 \includegraphics[width=0.7\linewidth]{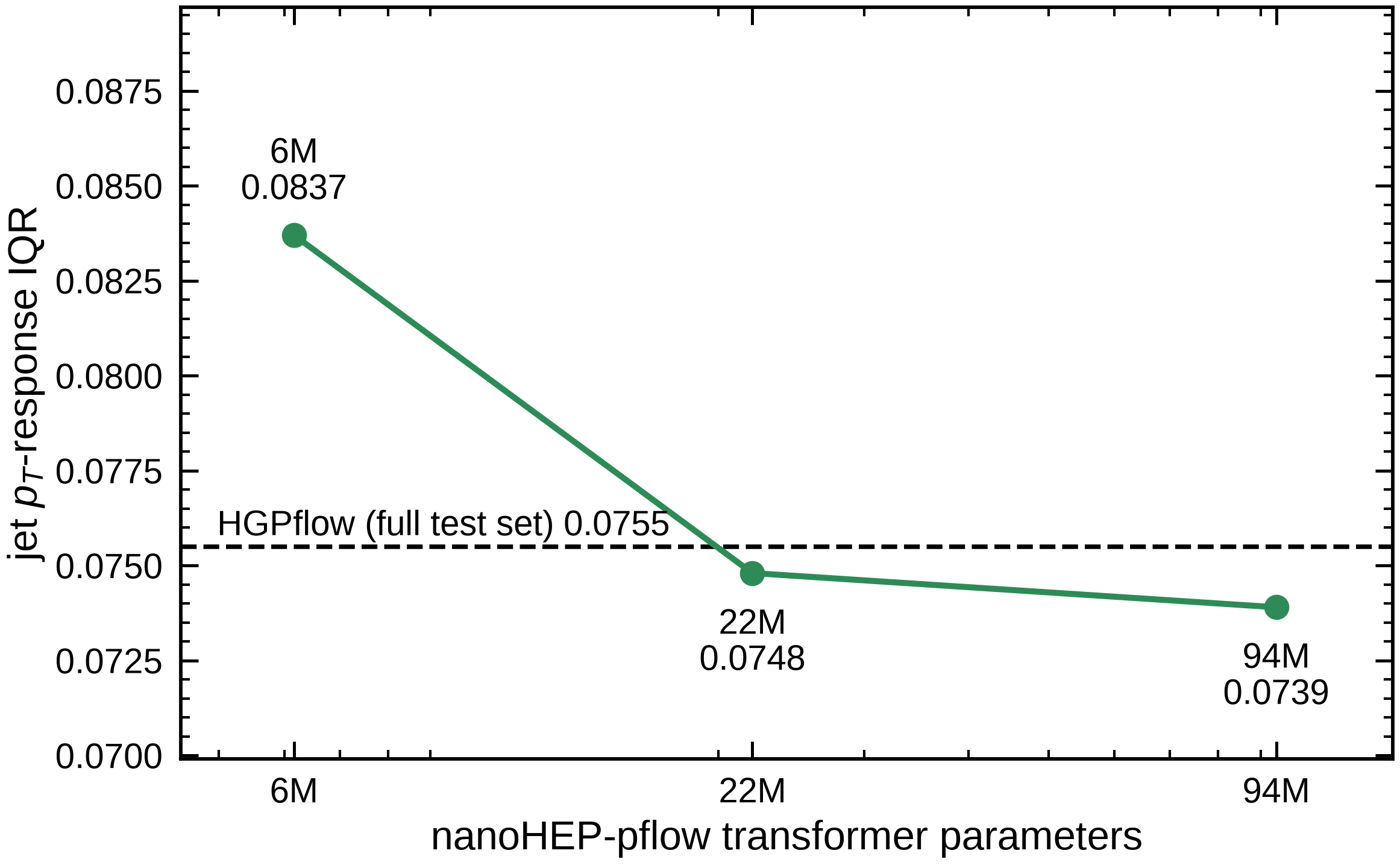}
 \caption{Jet $p_T$ response IQR versus parameter count for nanoHEP trained on particle flow alone, with HGPflow as the dashed line.}
 \label{fig:scaling_model}
\end{figure}

\begin{figure}[htbp]
 \centering
 \includegraphics[width=\linewidth]{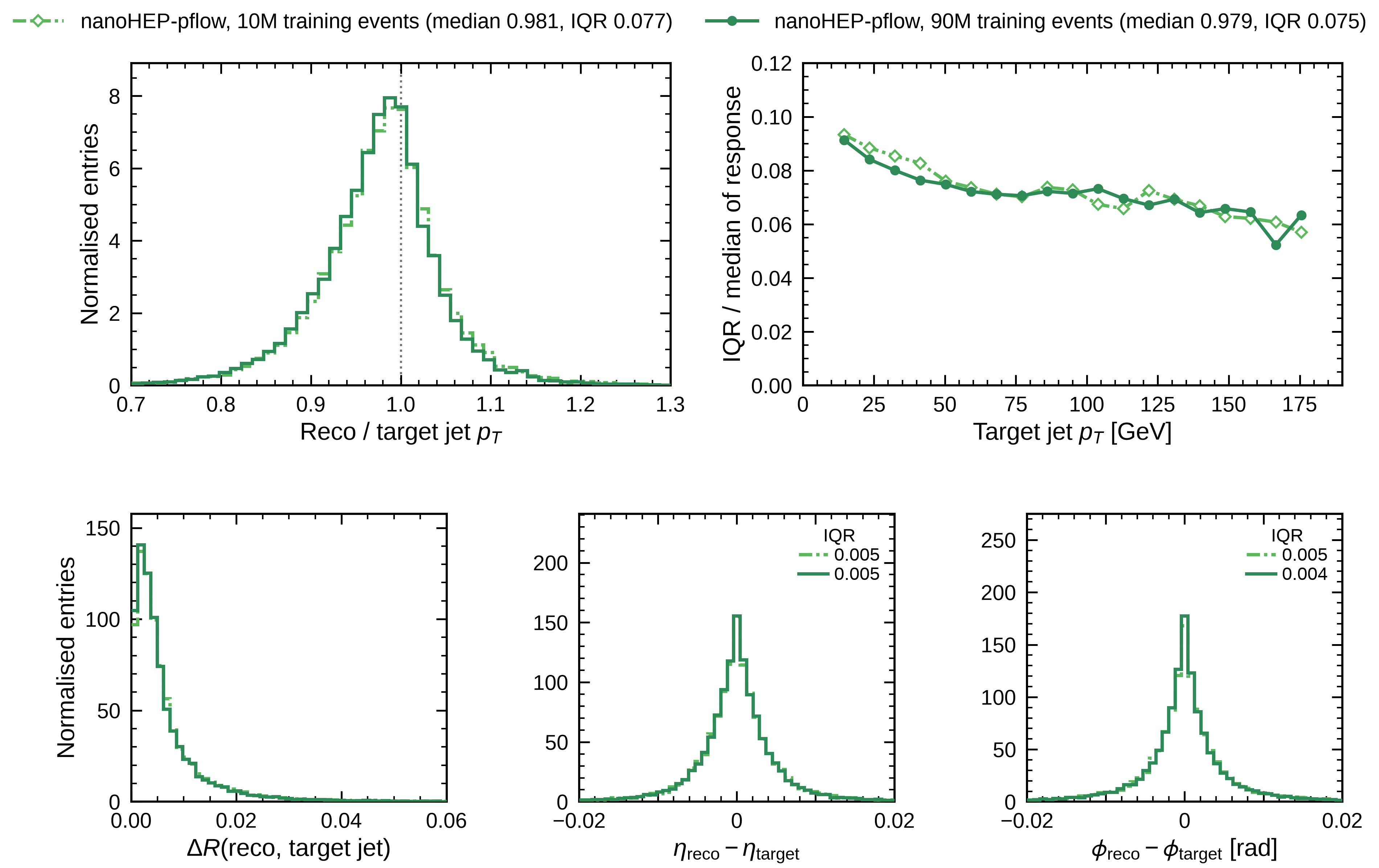}
 \caption{Jet $p_T$ response, response IQR/median versus target jet $p_T$, and the $\Delta R$, $\eta$ and $\phi$ residuals between reconstructed and target jets, for nanoHEP trained on particle flow alone with $10^7$ and $8.9\times10^7$ events.}
 \label{fig:scaling_data}
\end{figure}

We here present a simple study to indicate whether future work may benefit from a larger model or more data, with nanoHEP trained on particle flow alone. With parameter count, the jet $p_T$ response IQR is $0.084$ at $6$M parameters and $0.075$ at $22$M, with little further change at $94$M ($0.074$; Fig.~\ref{fig:scaling_model}). With training-set size, the IQR is $0.077$ with $10^7$ events and $0.075$ with $8.9\times10^7$ events (Fig.~\ref{fig:scaling_data}). The two runs also use different learning-rate schedules.

 \section{Particle-cloud measures of particle flow} \label{app:pflow_measures}

Table~\ref{tab:pflow_measures} collects every particle flow measure for every model, with argmax decoding. Besides the jet response and the classifier AUC of Sec.~\ref{sec:headline}, it gives two per-event measures.

The energy mover's distance (EMD)~\cite{Komiske2019EMD} is the minimum cost, in GeV, of moving the $p_T$ of the reconstructed particles onto the truth particles, with the cost per unit $p_T$ equal to the angular distance $\Delta R$. Since the two clouds carry different total $p_T$, we use the unbalanced form with $R=1$~\cite{Komiske2019EMD}. Each model is compared with the truth in its own output format: tokenised truth for nanoHEP, and raw truth for HEP4M and HGPflow. For nanoHEP, the tokeniser round trip alone contributes a median EMD of $0.17$~GeV, an order of magnitude below the model values.

The compatibility judge is a single classifier trained to separate an event's true pairing of inputs (tracks and topoclusters) and truth particles from a mismatched pairing. For the mismatched pairing, the event is the nearest neighbour in coarse event features. True and mismatched pairings are separated with an AUC of $0.98$. We report the mean logit of each model's reconstruction minus that of its own truth on the same events. A positive value means the reconstruction scores as more strongly bound to the inputs than the truth. The value is an ordering, not an accuracy, and its sign depends on the judge seed.

\begin{table}[htbp]
 \centering
 \small
 \setlength{\tabcolsep}{4pt}
 \begin{tabular}{lccccccc}
  \toprule
  & \multicolumn{2}{c}{jet $p_T$ response} & \multicolumn{2}{c}{EMD (GeV)} & classifier & \multicolumn{2}{c}{compatibility} \\
  \cmidrule(lr){2-3} \cmidrule(lr){4-5} \cmidrule(lr){7-8}
  model & median & IQR & median & vs HGPflow & AUC & seed 1 & seed 2 \\
  \midrule
  HGPflow          & $0.986$ & $0.0754$ & \textbf{2.09} & $0$ & $0.902$ & $+0.64$ & $+0.98$ \\
  nanoHEP-pflow    & $0.978$ & \textbf{0.0733} & \textbf{2.09} & $+0.024 \pm 0.003$ & \textbf{0.691} & $+0.53$ & $-0.31$ \\
  nanoHEP-multi    & $0.990$ & $0.0781$ & $2.16$ & $+0.048 \pm 0.003$ & $0.703$ & $+0.52$ & $-0.40$ \\
  HEP4M-pflow      & \textbf{0.994} & $0.0836$ & $2.62$ & $+0.31 \pm 0.01$ & $0.940$ & $+0.30$ & $-1.07$ \\
  HEP4M-multi      & $0.986$ & $0.0898$ & $2.78$ & $+0.46 \pm 0.01$ & $0.939$ & $+0.27$ & $-1.48$ \\
  \bottomrule
 \end{tabular}
 \caption{All particle flow measures, argmax decoding. Jet $p_T$ response median and IQR as in Fig.~\ref{fig:headline_jets}; EMD median over events and median per-event difference to HGPflow with its bootstrap standard error; classifier AUC as in Table~\ref{tab:pflow_main}; compatibility as the judge logit difference, for two judge seeds. Best value in the jet, EMD-median and AUC columns in bold.}
 \label{tab:pflow_measures}
\end{table}

On the jet response nanoHEP-pflow is narrower than HGPflow (paired IQR difference $-0.0021$, $95\%$ interval $[-0.0028, -0.0013]$), and the two are within $1\%$ on the EMD. The HEP4M models are $0.3$ to $0.5$~GeV behind on the EMD, in the same order as their jet resolution. HGPflow has the highest AUC of the HGPflow and nanoHEP rows, while its point estimates are as good as theirs.
 \section{Implementation details}
\label{app:implementation}

\subsection{Tokenisers} \label{app:modalities}

\begin{figure}[htbp]
 \centering
 \includegraphics[width=0.92\linewidth]{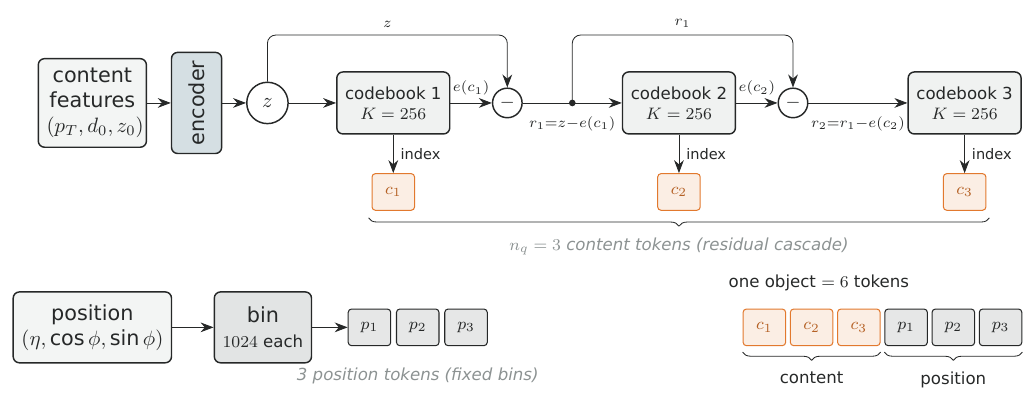}
 \caption{Residual VQ-VAE tokenisation of a single object. Top: the object's
 content features are encoded to one latent $z$ and quantised by a cascade of
 codebooks, each quantising the residual left by the previous, giving $n_q$
 content tokens. Bottom left: the position coordinates $(\eta,\cos\phi,\sin\phi)$
 are binned independently. Bottom right: the resulting per-object token block.}
 \label{fig:residual_vqvae}
\end{figure}

Each content tokeniser is a residual VQ-VAE~\cite{Lee2022RQVAE} with three codebooks of dimension $8$, around a small transformer encoder and decoder (Fig.~\ref{fig:residual_vqvae}). The codebook sizes are $256$ for tracks and topoclusters, $128$ for particles and $2048$ for cells. The jet tokeniser instead has eight parallel codebooks of size $128$ and dimension $16$. Features are power-transformed and min-max scaled before encoding. Each tokeniser is trained on its own modality, with a reconstruction and commitment loss and codebooks updated by exponential moving average, and is then frozen. Positions $(\eta, \cos\phi, \sin\phi)$ are binned uniformly, $1024$ bins each. We also tried a learned codebook for positions, with little benefit.

\subsection{Dataset} \label{app:dataset}

Each event holds one jet, whose $p_T$ is the vector sum of the truth-particle $p_T$. At most $256$ cells are stored per event.

\subsection{Cardinality head} \label{app:cardinality_head}

HEP4M predicts each modality's object count with a softmax over $0, \dots, N_{\max}$. In development, a regression head collapsed to a biased point estimate with no spread, and a per-slot indicator head gave no control over the total count. In a few rare directions the categorical head puts a spurious peak at zero, which argmax turns into an empty output, so we mask the zero bin at inference.

\subsection{Token factorisation}

Written out in tokens, with $k$ indexing the $n_q$ residual levels of an object's content code, Eq.~\ref{eq:parallel_factorisation} becomes
\begin{equation}
q_{\theta}(y, n \mid x) = \prod_{m} q_{\theta}(n_m \mid x) \prod_{i=1}^{n_m} \prod_{k=1}^{n_q} q_{\theta}\big(y^{m}_{i,k} \mid x, n, y^{m}_{i,<k}\big).
\end{equation}
In both models the residual levels of an object are sampled in order, each conditioned on the levels before it. Position tokens are sampled from their own heads. The two models differ only across objects: in nanoHEP each object is conditioned on the objects before it, and in HEP4M it is not.

\subsection{Training and inference} \label{subsec:training_inference}

The training settings are in Table~\ref{tab:training}. The nanoHEP loss is next-token cross-entropy on the output region only. The HEP4M loss is the sum of a per-codebook cross-entropy on the masked targets and the cardinality loss, with label smoothing and per-modality weights. Training is in bf16 and evaluation in fp32. At inference nanoHEP uses a key-value cache and a mask that keeps sampled sequences well-formed. We decode reconstruction with argmax and generation by sampling at $T=1$.

\begin{table}[htbp]
 \centering
 \small
 \begin{tabular}{lcc}
  \toprule
  & nanoHEP-multi & HEP4M-multi \\
  \midrule
  optimiser & AdamW~\cite{Loshchilov2019AdamW} & schedule-free AdamW~\cite{Defazio2024ScheduleFree} \\
  betas, weight decay & $0.9/0.95$, $0.1$ & $0.95/0.98$, $0.05$ \\
  peak LR, warmup & $1.2\times10^{-3}$, $2$k steps & $1.0\times10^{-3}$, $2.5$k steps \\
  schedule & cosine~\cite{Loshchilov2017SGDR} to $1.2\times10^{-4}$, $285$k steps & schedule-free \\
  effective batch & $4096$ & $16{,}384$ \\
  checkpoint & $296$k steps & lowest validation loss \\
  \bottomrule
 \end{tabular}
 \caption{Training settings of the multi-task models.}
 \label{tab:training}
\end{table}

\end{document}